\documentclass[12pt,dvips]{article}

\usepackage{rotating}
\usepackage{epsfig}
\usepackage{color}

\newcommand{\real}{\Re {\rm e}}

\newcommand{\lsim}{\raisebox{-0.13cm}{~\shortstack{$<$ \\[-0.07cm] $\sim$}}~}
\newcommand{\gsim}{\raisebox{-0.13cm}{~\shortstack{$>$ \\[-0.07cm] $\sim$}}~}

\newcommand{\sigmav}{\ensuremath{\langle\sigma v\rangle}}
\newcommand{\sigsip}{\ensuremath{\sigma^{\rm{SI}}_p}}

\newcommand{\mev}{\ensuremath{\,\mathrm{MeV}}}
\newcommand{\gev}{\ensuremath{\,\mathrm{GeV}}}
\newcommand{\tev}{\ensuremath{\,\mathrm{TeV}}}

\newcommand{\mx}{\ensuremath{m_{\chi^0_1}}}

\newcommand{\ohsq}{\ensuremath{\Omega_\chi h^2}}

\begin{document}

\def\thefootnote{\fnsymbol{footnote}}

{\small
\begin{flushright}
CNU-HEP-14-04, IPMU14-0344\\
\end{flushright} }

\medskip

\begin{center}
{\bf
{\Large
Dark Matter in Split SUSY with Intermediate Higgses} }
\end{center}

\smallskip

\begin{center}{\large
Kingman Cheung$^{a,b,c}$,
Ran Huo$^{d}$,
Jae Sik Lee$^{e}$,
Yue-Lin Sming Tsai$^{d}$
}
\end{center}

\begin{center}
{\em $^a$ Department of Physics, National Tsing Hua University,
Hsinchu 300, Taiwan}\\[0.2cm]
{\em $^b$ Division of Quantum Phases and Devices, School of Physics,
Konkuk University, Seoul 143-701, Republic of Korea}\\[0.2cm]
{\em $^c$  Physics Division, National Center for Theoretical Sciences, Hsinchu, Taiwan}\\[0.2cm]
{\em $^d$ Kavli IPMU (WPI), The University of Tokyo,
5-1-5 Kashiwanoha, Kashiwa, Chiba 277-8583, Japan}\\[0.2cm]
{\em $^e$ Department of Physics, Chonnam National University,
300 Yongbong-dong, Buk-gu, Gwangju, 500-757, Republic of Korea}\\[0.2cm]
(\today)
\end{center}

\bigskip

\centerline{\bf ABSTRACT}
The searches for heavy Higgs bosons and supersymmetric (SUSY) particles
at the LHC have left the
minimal supersymmetric standard model (MSSM) with an unusual spectrum
of SUSY particles, namely, all squarks are beyond a few TeV
while the Higgs bosons other than the one observed at 125 GeV could be 
relatively light. 
In light of this, we study a scenario characterized by two scales:
the SUSY breaking scale or the squark-mass scale $(M_S)$
and the heavy Higgs-boson mass scale $(M_A)$.
We perform a survey of the MSSM parameter space with $M_S \lsim 10^{10}$ GeV
and $M_A \lsim 10^4$ GeV such that the lightest Higgs boson mass is within
the range of the observed Higgs boson as well as satisfying a number of
constraints.
The set of constraints include
the invisible decay width of the $Z$ boson and that of the Higgs boson,
the chargino-mass limit, dark matter relic abundance from Planck,
the spin-independent cross section of direct detection by LUX,
and gamma-ray flux from dwarf spheroidal galaxies and gamma-ray line constraints
measured by Fermi LAT.
Survived regions of parameter space feature the dark matter with correct
relic abundance, which is achieved through either coannihilation with
charginos, $A/H$ funnels, or both.
We show that future measurements, e.g., XENON1T and LZ,
of spin-independent cross sections
can further squeeze the parameter space. \\

\vspace{0.3in}


\newpage

\section{Introduction}
Supersymmetry (SUSY) is one of the most elegant solutions, if not the
best, to the gauge hierarchy problem. SUSY provides an
efficient mechanism to break the electroweak symmetry dynamically
with a large top Yukawa coupling. Another
virtue is that the lightest SUSY particle (LSP) is automatically a dark
matter (DM) candidate to satisfy the relic DM abundance
assuming the $R$-parity  conservation.
The fine-tuning argument in the
gauge hierarchy problem requires SUSY particles at work at the TeV
scale to stabilize the gap between the electroweak scale and the grand
unified theory (GUT) scale or the Planck scale. 
With this scale the gauge coupling unification is also naturally 
achieved in renormalization group equation (RGE) running.

Although SUSY has quite a number of merits at least theoretically,
the biggest drawback of SUSY is that so far we have not observed any sign
of SUSY. Nevertheless, we have observed a light standard model (SM) like
Higgs boson,
which is often a natural prediction of SUSY. The null results for
all the searches of SUSY particles have pushed the mass scale of squarks
beyond a few TeV \cite{susy-exp}.
While abandoning SUSY as a solution to the gauge hierarchy problem, 
such a high-scale SUSY scenario also draws more and more attention on 
CP problems \cite{susycp}, cosmological problems \cite{susycosmo}, and 
DM search \cite{HSdm}.
On the other hand, the searches for the SUSY Higgs
bosons provide the less stringent mass limits and it still seems possible
to find them in the range of a few hundred GeV \cite{heavy-higgs}.
Consequently, we are left with an unusual spectrum of SUSY particles
and Higgs bosons: (i) all squarks are heavy beyond a few TeV \cite{susy-exp},
(ii) the gluino is heavier than about 1 TeV \cite{susy-gluino},
(iii) neutralinos and charginos can be of order $O(100-1000)$ GeV,
(iv) heavy Higgs bosons can be of order $O(100-1000)$ GeV \cite{heavy-higgs},
and (v) a light Higgs boson with a mass $125$ GeV \cite{atlas-cms}.
The spectrum is somewhat similar to the proposal of split SUSY
\cite{split}, except that
the heavy Higgs bosons need not be as heavy
as those of split SUSY.
We name the scenario the ``modified split SUSY'' framework,
with two distinct scales: the SUSY breaking scale $M_S$ and the heavy
Higgs-boson mass scale $M_A$.
In the following, for simplicity we call this ``modified split SUSY'' as
scenario A in which $M_S$ and $M_A$ are independent, while
the original split SUSY as scenario B in which $M_A$ and $M_S$ are set to be
equal. Since an extra TeV scale $M_A$ is obtained from cancellation of 
larger scales of $M_S$ or so, the fine tuning could be more serious than in the split SUSY.


We wish to be more specific and explicit about the framework and the motivation of our 
``modified split SUSY''.
In the MSSM, the mass of the lightest neutral Higgs boson
is basically determined by the weak gauge couplings and the
vacuum expectation values of the neutral components of the two Higgs doublets 
and, accordingly,
can not be much larger than the mass of the $Z$ boson.
While the mass scale of the other 4 Higgs states cannot be fixed by 
requiring the electrowek symmetry breaking. 
Usually, the arbitrary mass
parameter $M_A$ is introduced to fix the masses of the Higgs states
other than the lightest one. Our notion is that there is no compelling
reason for the scale $M_A$ to be equal to $M_S$ when we abandon SUSY
as a solution to the hierarchy problem.
With this choice of freedom we can modify the split SUSY (in split
SUSY $M_A = M_S$) to have two independent parameters $M_A$ and
$M_S$. With one more parameter, we can have more interesting collider
and dark matter phenomenology, as well as more viable regions of
parameter space, as we shall show in the main results.
We, therefore, come with an interesting variety of the
split SUSY. Instead of all the scalars being very heavy, we could have
the $M_A$ much lighter than $M_S$. This will have profound effects on
the dark matter phenomenology, especially the dark matter can
annihilation via the near-resonance of the heavy Higgs bosons. Since
the heavy Higgs bosons have much larger total decay widths than the
light Higgs bosons, the resonance effect of the Higgs boson would
enjoy much less fine tuning in giving the correct relic density of the
dark matter. Thus, interesting parameter space regions become viable
when $M_A$ goes down to sub-TeV and TeV ranges.

Phenomenologically, this modified split SUSY scenario is motivated by
the possibility that the Higgs bosons other than the one observed at
125 GeV can be relatively light compared to the high SUSY scale
$M_S$. If both $M_A$ and $M_S$ are set equal with $M_A < 10\tev$, as
will be shown in Fig. 1, 
only a small region with 
large $\tan \beta$ is allowed.
Nevertheless, if $M_A$ and $M_S$ are
set at different values, much larger parameter space with a wide range
of $\tan\beta$ will be allowed.
With more parameter space we can then contrast it with other existing
constraints. This is a strong motivation why we study this 
``modified split SUSY'' scenario. We can then perform a careful 
analysis using all dark matter constraints and collider limits.
%

In this work,
we consider the particle content of the minimal supersymmetric standard 
model (MSSM) in which the SUSY breaking scale or the sfermion-mass scale is denoted
by $M_S$.  The other scalar mass
scale is the mass of heavy Higgs bosons
characterized by $M_A$.
In split SUSY, all sfermions and heavy scalar
Higgs bosons are set a single scale $M_S$. However, in the modified split
SUSY scenario under consideration, $M_A$ can be substantially smaller 
than $M_S$. The gauginos and Higgsinos have masses in hundred GeVs and 
TeV. The lightest neutralino, the dark matter candidate,
will be composed of bino, wino, and Higgsino.
In addition to the neutralino-chargino coannihilation region, we 
also have the near-resonance regions of the $Z$ boson, the light Higgs boson,
as well as the heavy Higgs bosons, which is characterized by $M_A$.
It is the latter that makes the scenario different from the conventional
split SUSY. It is therefore important to explore this interesting 
scenario.

We perform a survey of the parameter space of the minimal
supersymmetric standard model (MSSM) characterized by two scales:
(i) the SUSY breaking scale $M_S$ with $M_S \lsim 10^{10}$ GeV, and
(ii) the heavy Higgs-boson mass scale $(M_A)$ with
$M_A \lsim 10^4$ GeV, such that
the lightest Higgs boson mass with large radiative corrections
from heavy squarks is within the range of the mass of
the observed Higgs boson.
We choose $M_A$ smaller than or at most equal to $M_S$.
Specifically, we assume the MSSM
above the SUSY breaking scale $M_S$.
Then we do the matching at the scale $M_S$ while we
decouple all the sfermions. We evolve from $M_S$ down to $M_A$ with
a set of RGEs
comprising of two-Higgs doublet model (2HDM),
gauge couplings, and gaugino couplings.
For this purpose, we derive the RGEs
governing the range between $M_S$ and $M_A$ and
present them in Appendix~\ref{sec:rge_ours}.
Then we do the matching at the scale $M_A$ while we decouple all the
heavy Higgs bosons. We evolve from $M_A$ down to the electroweak scale with
a set of RGE comprising of the SM and the gauginos. The matching is
then done at the electroweak scale. Once we obtain all the relevant
parameters at the electroweak scale, we calculate all the observables
and compare to experimental data.

In this work the LSP of
the MSSM is the DM candidate, which is the lightest
neutralino in the current scenario. 
Since we are strongly interested in DM, we include
a number of other existing constraints on SUSY particles and DM:
\begin{enumerate}
\item the invisible decay width of the $Z$ boson and that of the Higgs boson,
\item the chargino-mass limit,
\item dark matter relic abundance from Planck,
\item the spin-independent cross section of direct detection by LUX, and
\item gamma-ray flux from dwarf spheroidal galaxies (dSphs) and 
gamma-ray line constraints measured by Fermi LAT.
\end{enumerate}
%


Due to multidimensional model parameters involved in this work,
it will be advantageous to adopt a Monte Carlo sampling
technique to perform a global scan. In order to assess the robustness of 
our Monte Carlo results, 
we investigate both Bayesian maps in terms of marginal posterior (MP) and
frequentist ones in terms of the profile likelihood (PL) technique. 
However, the likelihood functions of experimental constraints are the same 
for both approaches.

The organization is as follows. In the next section, we describe the
theoretical framework of the modified split SUSY, including
the matching conditions at the scales of $M_S$ and $M_A$, and the
corresponding interactions of the particles involved. In Sec. 3, we list
the set of constraints from collider and dark matter experiments that
we use in this analysis. In Sec. 4, we present the results of our
analysis using the methods of PL and MP.
We discuss and conclude in Sec. 5.

\section{Theoretical Framework}

In the case under consideration,
we have the two characteristic scales:
the high SUSY scale $M_S$ and the Higgs mass scale $M_A$.
The relevant phenomenology
may be described by the effective Lagrangians depending on scale $Q$ as follows:
\begin{eqnarray}
M_S < Q \hphantom{\,< M_P}&: & {\cal L}={\cal L}_{\rm MSSM}
\nonumber \\
M_A < Q < M_S &:& {\cal L}={\cal L}_{\rm 2HDM}+{\cal L}^{(1)}_{\tilde\chi}
\nonumber \\
\hphantom{M_W <} Q < M_A &:& {\cal L}={\cal L}_{\rm SM}+{\cal L}^{(2)}_{\tilde\chi}
\end{eqnarray}

\subsection{Interactions for $M_A < Q < M_S$}

At the scale $M_S$ all the sfermions decouple when we assume that they are heavier
than or equal to the scale $M_S$. We are left with the spectrum of the
Higgs sector of the 2HDM, gauginos, and
higgsinos.

In this work, we take
the general 2HDM potential  as follows:
\begin{eqnarray}
\label{V2HDM}
\mathrm{V}_{\rm 2HDM} &=& -\mu_1^2 (\Phi_1^{\dagger} \Phi_1) - \mu_2^2
(\Phi_2^{\dagger} \Phi_2) - m_{12}^2 (\Phi_1^{\dagger} \Phi_2) -
m_{12}^{*2}(\Phi_2^{\dagger} \Phi_1) \nonumber \\
&&+ \lambda_1 (\Phi_1^{\dagger} \Phi_1)^2 +
\lambda_2 (\Phi_2^{\dagger} \Phi_2)^2 + 2\lambda_3 (\Phi_1^{\dagger}
\Phi_1)(\Phi_2^{\dagger} \Phi_2) + 2\lambda_4 (\Phi_1^{\dagger}
\Phi_2)(\Phi_2^{\dagger} \Phi_1) \nonumber \\
&&+ \lambda_5 (\Phi_1^{\dagger} \Phi_2)^2 +
\lambda_5^{*} (\Phi_2^{\dagger} \Phi_1)^2 + 2\lambda_6
(\Phi_1^{\dagger} \Phi_1) (\Phi_1^{\dagger} \Phi_2) + 2\lambda_6^{*}
(\Phi_1^{\dagger} \Phi_1)(\Phi_2^{\dagger} \Phi_1) \nonumber \\
&& + 2\lambda_7 (\Phi_2^{\dagger} \Phi_2) (\Phi_1^{\dagger} \Phi_2) +
2\lambda_7^{*} (\Phi_2^{\dagger} \Phi_2) (\Phi_2^{\dagger} \Phi_1)
\end{eqnarray}
with the parameterization
\begin{eqnarray}
\Phi_1 &=& -i\sigma_2\,H_d^*
= \left(\begin{array}{cc}
0 & -1 \\ 1 & 0
\end{array}\right)\,
\left(\begin{array}{c}
\frac{1}{\sqrt{2}}\,(v_d+H_d^0-iA_d^0) \\ -H_d^-
\end{array}\right)^*
= \left(\begin{array}{c}
H_d^+ \\ \frac{1}{\sqrt{2}}\,(v_d+H_d^0+iA_d^0)
\end{array}\right)\,; \ \ \
\nonumber \\
\Phi_2 &=& H_u = \left(\begin{array}{c}
H_u^+ \\ \frac{1}{\sqrt{2}}\,(v_u+H_u^0+iA_u^0)
\end{array}\right)\,; \ \ \
\end{eqnarray}
and $v_d=v \cos\beta=vc_\beta$, $v_u=v \sin\beta=vs_\beta$,
and $v\simeq 245$ GeV.
Then we have
\begin{eqnarray}
M_A^2&=&M_{H^\pm}^2+\lambda_4 v^2 -\real(\lambda_5)v^2\,, \\[3mm]
M_{H^\pm}^2&=&\frac{\real(m_{12}^2)}{c_\beta s_\beta}
-\frac{v^2}{c_\beta s_\beta}\left[\lambda_4 c_\beta s_\beta+
c_\beta s_\beta\real(\lambda_5)+
c_\beta^2\real(\lambda_6)+
s_\beta^2\real(\lambda_7) \right]\,,\nonumber
\end{eqnarray}
where $A=-s_\beta A_d^0+c_\beta A_u^0$ and
$H^+=-s_\beta H_d^+ +c_\beta H_u^+$.

The wino(bino)-Higgsino-Higgs interactions are given by
\begin{eqnarray}
{\cal L}_{\tilde\chi}^{(1)} & = &
\frac{1}{\sqrt{2}}H_u^\dagger\left(
\tilde{g}_{u} \sigma^a \widetilde{W}^a +
\tilde{g}_{u}^\prime  \widetilde{B} \right)\,\widetilde{H}_u
\nonumber \\
& + &
\frac{1}{\sqrt{2}}H_d^\dagger\left(
\tilde{g}_{d} \sigma^a \widetilde{W}^a -
\tilde{g}_{d}^\prime  \widetilde{B} \right)\,\widetilde{H}_d \ + \
{\rm h.c.}\,
\end{eqnarray}
where $\sigma^a$ are the Pauli matrices.
We note $H_d^\dagger=-\Phi_1^T\,i\sigma_2
=\left(\frac{1}{\sqrt{2}}\,(v_d+H_d^0+iA_d^0),-H_d^+\right)$.

\subsection{Matching at $M_S$}
The couplings of the interactions when $M_A<Q<M_S$ are determined by the
matching conditions at $M_S$ and the RGE evolution from $M_S$ to $Q$.
Assuming that all the sfermions are degenerate at $M_S$,
the quartic couplings at the scale $M_S$ are given by
\footnote{We neglect the stau contributions.}
\begin{eqnarray}
\lambda_1 &=&
\frac{1}{8}(g^2+g'^2)+\frac{N_c}{(4\pi)^2}\bigg(y_b^4\frac{A_b^2}{M_S^2}
(1-\frac{A_b^2}{12M_S^2})-y_t^4\frac{\mu^4}{12M_S^4}\bigg)\nonumber \\
\nonumber \\
\lambda_2 &=&
\frac{1}{8}(g^2+g'^2)+\frac{N_c}{(4\pi)^2}\bigg(y_t^4\frac{A_t^2}{M_S^2}
(1-\frac{A_t^2}{12M_S^2})-y_b^4\frac{\mu^4}{12M_S^4}\bigg)\nonumber \\
\lambda_3 &=&
\frac{1}{8}(g^2-g'^2)+\frac{N_c}{(4\pi)^2}\bigg(y_b^2y_t^2
\frac{A_{tb}}{2}+y_t^4(\frac{\mu^2}{4M_S^2}-\frac{\mu^2 A_t^2}{12M_S^4})
+y_b^4(\frac{\mu^2}{4M_S^2}-\frac{\mu^2 A_b^2}{12M_S^4})\bigg)\nonumber \\
\lambda_4 &=&
-\frac{1}{4}g^2+\frac{N_c}{(4\pi)^2}\bigg(-y_b^2y_t^2
\frac{A_{tb}}{2}+y_t^4(\frac{\mu^2}{4M_S^2}-\frac{\mu^2 A_t^2}{12M_S^4})
+y_b^4(\frac{\mu^2}{4M_S^2}-\frac{\mu^2 A_b^2}{12M_S^4})\bigg)\nonumber \\
\lambda_5 &=& -\frac{N_c}{(4\pi)^2}\bigg(y_t^4\frac{\mu^2
A_t^2}{12M_S^4}+y_b^4\frac{\mu^2 A_b^2}{12M_S^4}\bigg),
\nonumber \\
\lambda_6 &=& \frac{N_c}{(4\pi)^2}\bigg(y_b^4\frac{\mu
A_b}{M_S^2}(-\frac{1}{2}+\frac{A_b^2}{12M_S^2})+y_t^4\frac{\mu^3 A_t}{12M_S^4}\bigg),
\nonumber \\
\lambda_7 &=& \frac{N_c}{(4\pi)^2}\bigg(y_t^4\frac{\mu
A_t}{M_S^2}(-\frac{1}{2}+\frac{A_t^2}{12M_S^2})+y_b^4\frac{\mu^3 A_b}{12M_S^4}\bigg),
\label{eq:lam1to7}
\end{eqnarray}
with
\begin{equation}
A_{tb}=\frac{1}{6}\bigg(-\frac{6\mu^2}{M_S^2}-\frac{(\mu^2-A_bA_t)^2}{M_S^4}+\frac{3(A_b+A_t)^2}{M_S^2}\bigg)\,.
\end{equation}
We note that the quartic couplings at $M_S$ consist of
its tree level values and the threshold corrections induced by
the $A$ and $\mu$ terms. We further observe
$\lambda_{5,6,7}$ vanish without including the threshold corrections.

On the other hand, for the wino(bino)-Higgsino-Higgs couplings
at the scale $M_S$, we have
\begin{equation}
\tilde{g}^{(\prime)}_{u} = \tilde{g}^{(\prime)}_{d} = g^{(\prime)}\,.
\end{equation}
We note the relation $g^\prime = \sqrt{3/5}\,g_1$.

The threshold corrections to the gauge
and Yukawa couplings at $M_S$ also vanish
in the framework under consideration or
when all the sfermions are degenerate at $M_S$.

\subsection{Interactions for $Q < M_A$}
When the scale drops below $M_A$, all the heavy Higgs bosons decouple. We
are left with the SM particles, a light Higgs boson, gauginos, and
higgsinos.

The SM Higgs potential  is given by
\begin{equation}
V_{\rm SM}=\lambda \left[(\Phi^\dagger\Phi)^2-\frac{v^2}{2}\right]^2
\end{equation}
with
\begin{equation}
\Phi = \left(\begin{array}{c}
G^+ \\ \frac{1}{\sqrt{2}}\,(v+h+iG^0)
\end{array}\right)\,
\end{equation}
where $G^{\pm,0}$ denotes
the would-be Goldstone bosons and $h$ the physical neutral Higgs state.
We note $m_h^2=2\lambda v^2$.
The wino(bino)-Higgsino-Higgs interactions are then given by
\begin{eqnarray}
{\cal L}_{\tilde\chi}^{(2)} & = &
\frac{1}{\sqrt{2}}\Phi^\dagger\left(
\hat{g}_{u} \sigma^a \widetilde{W}^a +
\hat{g}_{u}^\prime  \widetilde{B} \right)\,\widetilde{H}_u
\nonumber \\
& + &
\frac{1}{\sqrt{2}}(-\Phi^T\,i\sigma_2)\left(
\hat{g}_{d} \sigma^a \widetilde{W}^a -
\hat{g}_{d}^\prime  \widetilde{B} \right)\,\widetilde{H}_d \ + \
{\rm h.c.}\,.
\end{eqnarray}

\subsection{Matching at $M_A$}
The couplings of the interactions when $Q<M_A$ are determined by the
matching conditions at $M_A$ and the RGE evolution from $M_A$ to $Q$.

At the scale $M_A$,
the quartic coupling $\lambda$ of the SM Higgs potential
is given by
\begin{eqnarray}
\lambda &=& \lambda_1\cos^4\beta + \lambda_2\sin^4\beta + 2\tilde{\lambda}_3\cos^2\beta
\sin^2\beta
\nonumber \\
&+& 4\lambda_6\cos^3\beta\sin\beta + 4\lambda_7\cos\beta\sin^3\beta
+ \delta\lambda
\label{eq:lambda_MA}
\end{eqnarray}
where $\tilde{\lambda}_3=\lambda_3+\lambda_4+\lambda_5$ and
$\delta\lambda$ denotes the threshold correction.
We find that the threshold correction to $\lambda$ is given by
\begin{eqnarray}
\delta\lambda&=&\frac{1}{4\pi^2}\bigg[
\left(\lambda_A^3\frac{v^2}{M_A^2}-\frac{1}{3}\lambda_A^4\frac{v^4}{M_A^4}
\right) +
\left(\lambda_H^3\frac{v^2}{M_H^2}-\frac{1}{3}\lambda_H^4\frac{v^4}{M_H^4}
\right) +
2\left(\lambda_\pm^3\frac{v^2}{M_{H^\pm}^2}-\frac{1}{3}\lambda_\pm^4\frac{v^4}
{M_{H^\pm}^4}\right)\bigg] \nonumber \\[2mm]
&+&
\frac{1}{8\pi^2}\left( \lambda_H^2\ln\frac{M_H}{M_A} +
2\lambda_\pm^2\ln\frac{M_{H^\pm}}{M_A} \right)
\end{eqnarray}
where $M_H$ denotes the mass of the heavier CP-even neutral Higgs boson
and the couplings $\lambda_{A,H,\pm}$ are defined as follows:
\begin{eqnarray}
\lambda_A&=&(\lambda_3+\lambda_4-\lambda_5) +
(\lambda_1+\lambda_2-2\tilde\lambda_3)\cos^2\beta\sin^2\beta
+(-\lambda_6+\lambda_7)\sin2\beta\cos2\beta\,,\nonumber\\[2mm]
\lambda_H&=&\tilde\lambda_3 +3(\lambda_1+\lambda_2-2\tilde\lambda_3)
\cos^2\beta\sin^2\beta + 3(-\lambda_6+\lambda_7)\sin2\beta\cos2\beta
\,,\nonumber\\[2mm]
\lambda_\pm&=&\lambda_3+(\lambda_1+\lambda_2-2\tilde\lambda_3)\cos^2\beta\sin^2\beta
+ (-\lambda_6+\lambda_7)\sin2\beta\cos2\beta.
\end{eqnarray}
The wino(bino)-Higgsino-Higgs couplings at $M_A$ are given by
\begin{equation}
\hat{g}^{(\prime)}_{u} = \tilde{g}^{(\prime)}\,\sin\beta\,; \ \ \
\hat{g}^{(\prime)}_{d} = \tilde{g}^{(\prime)}\,\cos\beta\,.
\end{equation}
The threshold corrections to the gauge
and Yukawa couplings at $M_A$ are neglected  because of
the approximated degeneracy among $M_A$, $M_H$, and $M_{H^\pm}$.

\subsection{Matching at the electroweak scale}
Matching at the electroweak scale is exactly the same as in
the original split SUSY framework. We closely follow Ref.~\cite{Giudice:2011cg}
to include the threshold corrections to
the gauge couplings at the electroweak scale
and to calculate the pole masses for the Higgs boson and the top quark.

Since we are adopting the one-loop matching conditions, see Eqs.~(\ref{eq:lam1to7}) and (\ref{eq:lambda_MA}), 
it is more appropriate to employ two-loop RGEs.  
However, not all the two-loop RGEs are available for the present framework, 
and the higher-order corrections may be minimized by the judicious choice of the top-quark mass 
for the scale where the lightest Higgs mass is estimated. Our approach is to be considered as 
an intermediate step towards the more precise calculation of the lightest Higgs mass in our modified split SUSY scenario.

\section{Experimental Constraints and Likelihoods}
In this section, we describe how to
construct the likelihood functions
involved with experimental constraints which are used in both MP and PL 
approaches.
For the experimental constraints considered in this work, we assume
either half-Gaussian or Gaussian distribution
when the central values $\mu$, experimental errors $\sigma$,
and theoretical errors $\tau$ are available.
Otherwise, we take Poisson distributions.

In Table~\ref{tab:exp_constraints}, in the second last column, we show the
likelihoods of each experimental constraint. Here ``hard cut" means we
apply the 95\% upper limits instead of constructing its likelihood.
For the details of our statistical treatment,
we refer to Appendix \ref{sec:stat}.
In the following subsections, we give more
details of the constraint and likelihood
of each measurement.

\begin{table}[t]\footnotesize
\begin{center}
\begin{tabular}{|l|l|l|l|l|l|}
\hline 
Measurement & central value $\mu$ & Error:~($\sigma$,~$\tau$) & Distribution & Ref.\\
\hline
$\Gamma_{\rm{inv}}^Z$ & $499\,{\rm MeV}$ & $1.5\,{\rm MeV}, 0.0 $ & Gaussian
& \cite{Beringer:1900zz} \\
$\Gamma_{\rm{inv}}^h$ & $0.1\,{\rm MeV}$ & $^{+0.51}_{-0.41}\,{\rm MeV}, 0.0 $ & Gaussian
& \cite{Cheung:2013kla} \\
$m_{\chi^\pm}$ & $103.5\,{\rm GeV}$ & $0.0\,{\rm GeV}, 1\% $ & half-Gaussian &\cite{LEP} \\
\hline
relic abundance                   & $0.1186$      & $0.0031$,~$10\%$              & half-Gaussian
&  \cite{Ade:2013zuv}\\
\hline
LUX (2013)                      & see Ref. \cite{Matsumoto:2014rxa}
& see Ref. \cite{Matsumoto:2014rxa}              & Poisson &
\cite{Akerib:2013tjd}\\
\hline
dSphs $\gamma$-ray                      & see Ref. \cite{Tsai:2012cs}
& see Ref. \cite{Tsai:2012cs}              & Poisson & \cite{Ackermann:2013yva}\\
Monochromatic $Z^0\gamma$ and $\gamma\gamma$     & 95\% upper limits    & 95\% upper limits
& hard-cut & \cite{Ackermann:2013uma}\\
\hline 
\end{tabular}\caption{The experimental constraints and
the likelihoods. Where it is applicable,
the central value $\mu$, experimental error $\sigma$,
and theoretical error $\tau$ are given.
}
\label{tab:exp_constraints}
\end{center}
\end{table}

\subsection {Colliders}

\subsubsection{Invisible decay widths}


The invisible decay width of the $Z$ boson was accurately measured by
taking the difference between the total width and the visible width,
and is well explained by the three light active neutrino species of the SM.
Any additional invisible decays of the $Z$ boson are strongly
constrained by this data. In the current framework, the additional
invisible width comes from $Z\to\chi^0_1\chi^0_1$.
With the invisible width given in the PDG~\cite{Beringer:1900zz},
$ \Gamma^Z_{inv}  = 499 \pm 1.5$ MeV, we can constrain $Z\to\chi^0_1\chi^0_1$.

If the neutralino mass is below $m_h/2$, the Higgs boson can decay into
a pair of neutralinos, thus contributing to an invisible width of the
Higgs boson. From a global fit using the Higgs-boson data at the
7 and 8 TeV runs of the LHC, the invisible width of the Higgs boson
is constrained to be $\Gamma^h_{\rm inv} < 0.6$ MeV~\cite{Cheung:2013kla}
at 1-$\sigma$ level if all other parameters are fixed at their SM values.
If other parameters are allowed to vary, the $\Gamma^h_{\rm inv}$ would
have a more relaxed limit, which is about the same as
the bound from the direct search on the
invisible mode of the Higgs boson, which has
a branching ratio about $50\%$~\cite{invisible}.
Nevertheless, we use $\Gamma^h_{\rm inv} < 0.6$ MeV in this work,
as shown in Table~\ref{tab:exp_constraints}.

\subsubsection{Chargino mass}
The mass limits on charginos come either from direct search or
indirectly from the constraint set by the non-observation of
${\chi}^0_2$ states on the gaugino and higgsino MSSM parameters
$M_2$ and $\mu$. For generic values of the MSSM parameters,
limits from high-energy $e^+ e^-$ collisions coincide with the highest
value of the mass allowed by phase space, namely $m_{{\chi}^\pm} \lsim
 \sqrt{s}/2$.
The combination of the results of the four LEP collaborations
of LEP2 running at $\sqrt{s}$ up to 209 GeV yields a
lower mass limit of $m_{{\chi}^\pm_1} \ge 103.5$ GeV, which
is valid for general MSSM models. However, it could be weakened
in certain regions of the MSSM parameter space
where the detection efficiencies or production cross sections are
suppressed, e.g., when the mass difference
$m_{{\chi}^\pm_1} -m_{{\chi}^0_1}$ becomes too small.
Regardlessly, we simply employ the mass limit of
$m_{{\chi}^\pm_1} \ge 103.5$ in this work.
We do not use the LHC constraint  since it is
more model dependent and does not give any bounds when $m_{{\chi}^0_1}
  \gsim 70$ GeV~\cite{Chatrchyan:2012pka}.
Furthermore, for $m_{\chi^0_1}< 70\gev$ region, the $H/Z$ resonance region
(see next subsection) is not sensitive to this search~\cite{Han:2014sya}.
Note that the $\chi\chi^\pm$ coannihilation is strongly forbidden
by this limit especially when $\mx\lsim 90\gev$.

To deal with the chargino mass limit without detector simulations,
we adopt the half Gaussian distribution
when $m_{\chi_1^\pm}<103.5\gev$ to describe the tail
of the chargino mass likelihood function.
For the likelihood, we assume $\sim 1\%$ theoretical uncertainty.
When $m_{\chi_1^\pm}\geq 103.5\gev$,
we always assume the maximum likelihood.


\subsection{Relic abundance}

The half-Gaussian distribution for relic abundance likelihood in
Table~\ref{tab:exp_constraints} suits the well-motivated moduli
decay scenario~\cite{Choi:2005uz,Conlon:2006us,Conlon:2006wz,Acharya:2007rc,Acharya:2008zi,Fan:2013faa,Blinov:2014nla}.
In this scenario, the relic abundance can be reproduced by moduli
decay after the freeze-out, which is different from the usual 
multi-component DM scenario,
in which the total relic abundance is shared among a few DM candidates, such as the axion.
In the moduli decay scenario, all the DM is still assumed to be the neutralino, 
and the DM local density need not be rescaled
with respect to the neutralino fraction as implemented in 
the multi-component DM scenario,
so that the DM direct and indirect detection constraints will be stronger.


Very often, the neutralino DM in most of the MSSM parameter space
over-produces the relic abundance, 
because the annihilation in the early Universe is too inefficient.
Generally speaking, by opening the $W^+W^-$ final state
the wino-like neutralino can very efficiently reduce relic abundance
for wino mass up to $3-4\tev$,
e.g. see Ref.~\cite{Fan:2013faa,Hisano:2006nn, Mohanty:2010es}.
However, it requires some specific mechanisms for bino-like, Higgsino-like, 
or mixed neutralinos to fulfill correct relic abundance.
Sometimes more than one mechanisms are needed.
In most cases the (non-wino) regions both of correct relic abundance and 
still allowed by the
current LHC direct searches in our modified split SUSY parameter space are:

\begin{itemize}
\item The $Z/h$ resonance region, where the neutralinos annihilate through
the resonance with the
$Z$ boson at $\mx\sim 45\gev$ and Higgs boson at $\mx\sim 62.5\gev$.
In this region, neutralinos are governed mainly by the bino fraction but
with a small mixing with the higgsino fraction.

\item The chargino-neutralino coannihilation region, where the $\mu$ parameter
is usually closed to gaugino parameters $M_1$ or $M_2$
so that the $\chi^0_1$, $\chi^\pm_1$, and
$\chi^0_2$ are almost degenerate.
If the masses between $\chi^0_1$ and $\chi^\pm_1$
or $\chi^0_1$ and $\chi^0_2$ are very close to each other,
the number densities of the next-to-lightest supersymmetric particle(s)
(NLSP(s)) have only slight Boltzmann suppression with respect to the
LSP number density.
Therefore, all the interactions among the LSP and NLSP(s),
such as $\chi^0_1-\chi^\pm_1$,
$\chi^0_1-\chi^0_2$ and $\chi^\pm_1-\chi^0_2$, play important
roles to reduce the relic abundance.
Note that $\chi^0_1$ in this region shall have nonnegligible fractions of
wino or higgsino in order to coannihilate with $\chi^\pm_1$ and $\chi^0_2$.

\item The $A/H$ funnel region, where neutralinos annihilate through
  the resonance of the pseudoscalar Higgs boson $A$ or the heavy
  scalar Higgs boson $H$. In the original split SUSY framework with
  $M_A=M_S$, because of the large mass of $A/H$ as well as
  their large decay width, this mechanism becomes irrelevant.  
  On the other hand, in our
  modified split SUSY scenario with light $M_A$, this $A/H$
  funnel can still play a significant role in reducing the relic
  abundance. Nevertheless, we shall see later that the $A/H$-funnel for
  $\mx>1\tev$ is not efficient enough to reduce the relic abundance
  because of the larger $A/H$ decay width.

\end{itemize}

In split SUSY scenario, because of the very heavy sfermion masses, all
the $\tilde{f}-\chi$ coannihilation channels have been closed.  On the other
hand, the chargino annihilation is still allowed but the chargino mass must be
above the LEP limit, $m_{\chi^\pm}>103.5$ GeV.
We found in our viable parameter space
the majority of bino-like neutralino and chargino is always close to
each other ($\chi\chi^\pm$ coannihilation on).
Besides, $\chi\chi$ annihilation can have a few other choices.
Lowering $M_A$ to less than 1\tev, the $A/H$-funnel region can be important,
especially for higgsino and mixed neutralino.  For $\mx<100\gev$, $Z$-
and $h$- resonances can also significantly reduce relic abundance.
Finally, the wino-like neutralinos can easily annihilate into the
$W^+W^-$ final state, which can sufficiently reduce relic abundance as well.

\subsection{LUX: spin-independent cross section}

At present the most stringent 90\% C.L. limit on the spin-independent 
component of the
elastic scattering cross section comes from LUX~\cite{Akerib:2013tjd}.
However, it did not take into account the systematic uncertainties
from nuclear physics and astrophysics,
otherwise the constraint becomes much less straightforward.
The astrophysical uncertainties mainly come from our poor knowledge of
the DM local density and velocity distribution.  In order to account for 
the uncertainties of all the astrophysical parameters, we
adopt the phase-space density factor and its associated error bars as
computed in Ref.~\cite{Catena:2011kv}.
Nuclear physics uncertainties enter the systematic uncertainties through
the nuclear matrix elements, mainly the pion-nucleon
sigma term $\Sigma_{\pi N}$ and
the strange quark content of the nucleon $f_{Ts}$,
which promote the spin-independent cross sections from quark level into 
nucleon level.
In Table~\ref{fig:inputs},
we treat the  $\Sigma_{\pi N}$ and $f_{Ts}$
as nuisance parameters and distribute as Gaussian with
central values and error bars obtained by recent lattice QCD calculations.
Regarding the reconstruction of the LUX likelihood including the
astrophysical and nuclear uncertainties, we refer to
Ref.~\cite{Matsumoto:2014rxa} for more detailed explanations.

\subsection{Fermi LAT gamma ray }
\subsubsection{Continuous gamma ray from dSphs}

The most luminous gamma-ray source is the Galactic Center (GC) in the Milky Way,
but it is also subject to higher astrophysical backgrounds.
Better constraints were obtained from the diffuse gamma rays
from the dSphs of the Milky Way.
They are less luminous and dominated by DM, with little 
presence of gas or stars.
Recently, the Fermi LAT Collaboration improved significantly the
previous sensitivities to DM searches from dSphs~\cite{Ackermann:2013yva}.

Unlike the published limit from the Fermi LAT collaboration,
we only include the eight classical dSphs in our analysis,
because the DM halo distribution in the classical dSphs is measured with a
higher accuracy from the velocity dispersion of the luminous
matter~\cite{Martinez:2009jh}.
We use the 273 weeks’ Fermi-LAT data and the Pass-7 photon selection
criteria, as implemented in the \texttt{FermiTools}.
The energy range of photons is chosen from $200\mev$ to $500\gev$,
and the region-of-interest is adopted to be
a $14^\circ\times 14^\circ$ box centered on each dSphs.
The J-factors are taken from Table-I in Ref.~\cite{Ackermann:2013yva}.

In the likelihood analysis, the Fermi-LAT data are binned into
11 energy bins logarithmically spaced between 0.2 and 500 GeV, and we
calculate the
likelihood map of Fermi-LAT dSphs on the Ebin-flux plane following
the method developed in Ref.~\cite{Tsai:2012cs}.

\subsubsection{Fermi photon line measured from GC }

The experimental signature of monochromatic lines 
over the continuous spectrum is a clean signal of DM annihilation. 
In MSSM, the annihilation of $\chi^0_1\chi^0_1$ into photons 
induced by loop diagrams
also provides stringent constraints on parameter space, especially when
$\chi^0_1$ is wino-like and the annihilation cross section is enhanced.
However, we do not reconstruct the likelihood for the Fermi-LAT photon line
experiment but simply take the published limit at $5\gev<\mx<300\gev$.
In addition, we adopted the Isothermal profile since it is known
to be more conservative than NFW or Einasto profile~\cite{Ackermann:2013uma}.

%
%
\section{Numerical Analysis}
%
%

In this section,
after describing the input parameters over which we
perform the scan of the MSSM,
we present the results of our numerical study.
%

To compute the DM observables such as
the relic abundance $\ohsq$, DM-proton elastic scattering
cross section $\sigsip$, annihilation cross section $\sigmav$ at the present
time, and branching ratios of DM annihilation,
we calculate couplings and
mass spectra at the neutralino-mass scale $M_\chi\equiv\sqrt{\mu \times M_2}$,
where $\mu$ and $M_2$ denote the values at the scale $M_\chi$.
First, we solve the RGEs from $M_S$ to $M_A$ with those given in
Appendix~\ref{sec:rge_ours}.
For the evolution from $M_A$ to $M_\chi$, which is required
when $M_\chi<M_A$, we employ the split SUSY RGE code
\footnote{We thank Pietro Slavich for providing us the \texttt{SplitSuSpect}
code~\cite{Bernal:2007uv}.}.
Then we generate the \texttt{SLHA} output and
feed it into \texttt{DarkSUSY~5.1.1}~\cite{Gondolo:2004sc} to
compute the DM observables.  Finally,
we use the DM annihilation information from \texttt{DarkSUSY~5.1.1}
to compute the likelihoods for direct and indirect detections by following the
method developed in Ref.~\cite{Matsumoto:2014rxa}.

We perform the MSSM parameter space scan, including nuisance parameters,
by use of \texttt{MultiNest v2.18} \cite{Feroz:2008xx} taking $15,000$ living
points with a stop tolerance factor of $0.01$ and an enlargement factor
of $0.8$.

\subsection{Input Parameters}
In this subsection,
we provide detailed description of our MSSM input parameters
and the nuisance parameters. For the SM input parameters we take 
the PDG values \cite{Beringer:1900zz}.

%
\begin{table}[t]
\begin{center}
\begin{tabular}{|c|c|c|}
\hline
\hline
MSSM Parameter & Range & Prior distribution\\
\hline
bino mass&  $10^{-2}<|M_1|/{\rm TeV}<5$ & Log\\
wino mass& $9\times 10^{-2}<|M_2|/{\rm TeV}<5$ & Log\\
$\mu$ & $9\times10^{-2}<\mu/{\rm TeV}<5$ & Log\\
gluino mass& $1<|M_3|/{\rm TeV} <5$ & Log\\
$\tan\beta$ & $2<\tan\beta <62$ & Flat\\
\hline
$M_A$& $0.2\,{\rm TeV}<M_A<\rm{min}\,[10\,{\rm TeV},M_S]$ & Flat (Scenario A)\\
 &  $M_A=M_S$& Fixed (Scenario B)\\
\hline
\hline
Nuisance Parameter & Central value and systematic uncertainty & Prior distribution\\
\hline
$m_{h}$ ($\gev$) & $125.1\pm 2.0$~\cite{Aad:2014aba,Khachatryan:2014ira} &
Gaussian \\
$\Sigma_{\pi N}$ (\mev) & $41.0\pm 6.4$~\cite{Alvarez-Ruso:2013fza} & Gaussian\\
$f_{Ts}$ & $0.043\pm 0.011$~\cite{Junnarkar:2013ac} & Gaussian\\
\hline
\hline
\end{tabular}
\caption{The prior ranges and distributions of the
input parameters over which we perform the
scan of the MSSM.
}
\label{tab:inputs}
\end{center}
\end{table}

In Table \ref{tab:inputs}, the input parameters,
their prior ranges and types of prior distributions are shown.
We take $|M_{1,2}|\,,|\mu|\,<5$ TeV because
it is hard to satisfy the relic abundance constraint
with the LSP heavier than $3-4$ TeV.
We apply the same maximum value for the
gluino mass parameter, which does not affect our results much.
The smallest values of $|M_{2}|$ and $\mu$ are chosen by taking into account
the LEP limit on the chargino mass. We are taking
$|M_{3}|>1$ TeV because of the LHC limit on the gluino mass.
We cover the range of $\tan\beta$ up to $62$
and fix the trilinear parameter $A_0=\mu\cot\beta$
assuming the no-mixing scenario in the stop sector.
The MSSM input parameters $M_{1,2,3}$, $\mu$, and $A_0$ are given
at the scale $M_S$
while $\tan\beta$ is the value at the scale $M_A$.

Note that, in this work, we are using $m_h$ as an input
nuisance parameter and, accordingly, the value
of the high SUSY scale $M_S$ is an output.
Numerically, we solve the RGEs to find the value of
$M_S$ which gives the input value of $m_h$.
The Higgs boson mass measurements in the diphoton decay channel now give
$m_h=125.4 \pm 0.4$ GeV (ATLAS) \cite{atlas_aa_2014} and
$m_h=124.70 \pm 0.31~({\rm stat}) \pm 0.15~({\rm syst})$ GeV
(CMS)~\cite{david}.
On the other hand, the theoretical error of Higgs mass
is estimated to be around $2-3\gev$~\cite{Heinemeyer:2011aa} which is much
larger than the experimental errors of $\sim 0.4$ GeV.
Therefore, in this work, we  are taking
$m_h= 125.1\gev$ with a Gaussian experimental uncertainty
of $\sigma=2\gev$.

Depending on the relative size of $M_A$ to $M_S$,
we are taking two scenarios:
\begin{itemize}
\item scenario A: $M_A\leq\rm{min}[10\tev, M_S]$, 
\item scenario B: $M_A=M_S$  (the same as the original split SUSY).
\end{itemize}
In the scenario A, we are taking the maximum value of 10 TeV for $M_A$,
because the $A/H$-funnel ($M_A \sim 2\,m_{\chi^0_1}$) mechanism
becomes ineffective for neutralino annihilation when $M_A$ is beyond 10 TeV.
Smaller values of $M_A$ may help to obtain the correct Higgs-boson mass
when $M_S$ is too large to give $m_h\sim 125$ GeV in
the original split SUSY framework.
On the other hand, the choice of $M_A$ in scenario B is the same as in
the original split SUSY framework.
We note that the scenario B is a part of scenario A
if $M_S<10\tev$.

We further need inputs for the pion-nucleon sigma term $\Sigma_{\pi N}$
and the strange quark content of the nucleon $f_{Ts}$.
To account for the systematic uncertainties involved in the evaluation
of the relevant nuclear matrix elements,
we also treat them as nuisance parameters, as mentioned before.
The central values and errors are obtained by recent lattice QCD calculations.

\subsection{Numerical Results}

\begin{figure}[t!]
\centering
\includegraphics[width=5in]{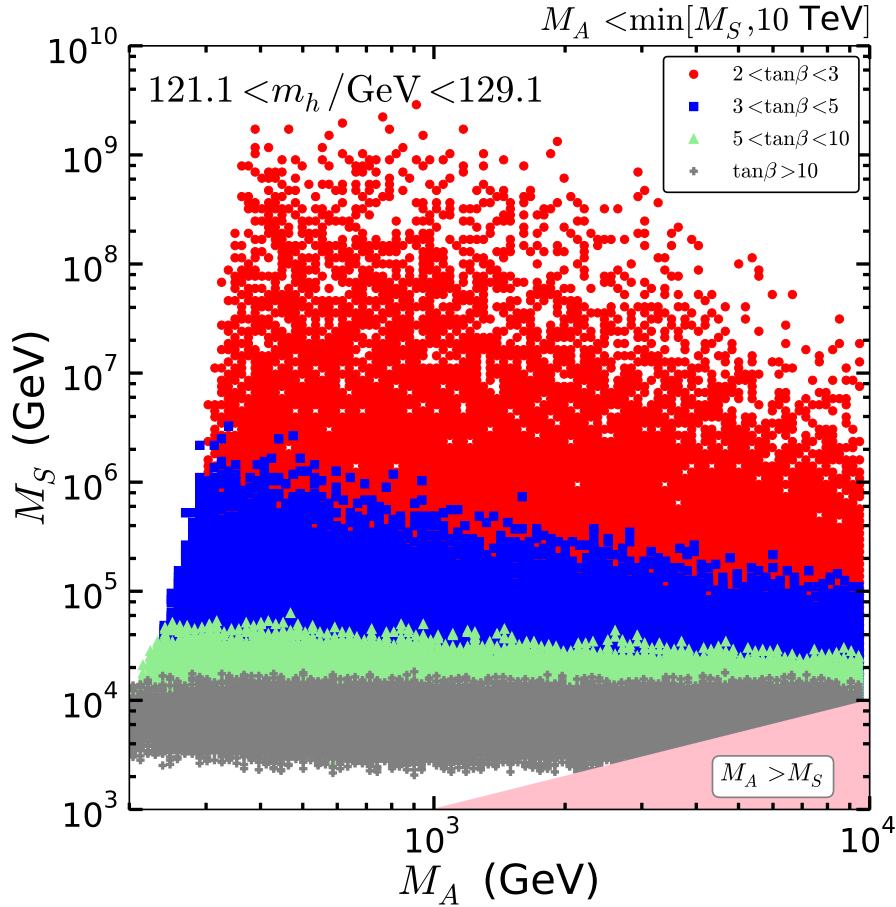}
\caption{\label{fig:mA_mS}
The scatter plot on the ($M_A$, $M_S$) plane varying input parameters
as in Table~\ref{tab:inputs} while requiring $m_h$ to be in the
$2$-$\sigma$ range: $121.1<m_h/\gev<129.1$.
The color scheme are: $2<\tan\beta<3$ (red circle),
$3<\tan\beta<5$ (blue square),
$5<\tan\beta<10$ (green triangle), and
$\tan\beta<10$ (gray cross).
In the pink region, $M_A>M_S$ which is
out of our current consideration.
}
\end{figure}

We are taking both the PL and MP methods
and make comparisons where it is informative.
We note that,
when we present our result based on the MP method,
the systematic uncertainties of the input parameters
are automatically included by utilizing a Gaussian prior distribution,
see the nuisance parameters in Table \ref{tab:inputs}.
On the other hand, when we are using the PL method,
the systematic uncertainties are added to the likelihood function.

In Fig.~\ref{fig:mA_mS} we show
the scatter plot on the ($M_A$, $M_S$) plane by varying input parameters
as in Table~\ref{tab:inputs},
while requiring $m_h$ to be in the $2$-$\sigma$ range:
 $121.1\gev<m_h<129.1\gev$.
Different colors represent different $\tan\beta$ ranges.
We observe that a larger $M_S$ is required for
small values of $\tan\beta$ and also as $M_A$ decreases.
When $\tan\beta\gsim 10$, $M_S$ becomes almost
independent of $M_A$ and it lies between $\sim 3$ TeV and $\sim 15$ TeV.
When $M_A=M_S$ is taken as in the scenario B, 
the value of $M_S$ is smaller in order to achieve $m_h\sim125$ GeV.
Therefore, in the split-SUSY framework with the intermediate Higgses
lighter than $\sim 10$ TeV, $M_S$ is generally predicted to be higher
especially when $\tan\beta$ is small.

\begin{figure}[t!]
\centering
\includegraphics[width=3.0in]{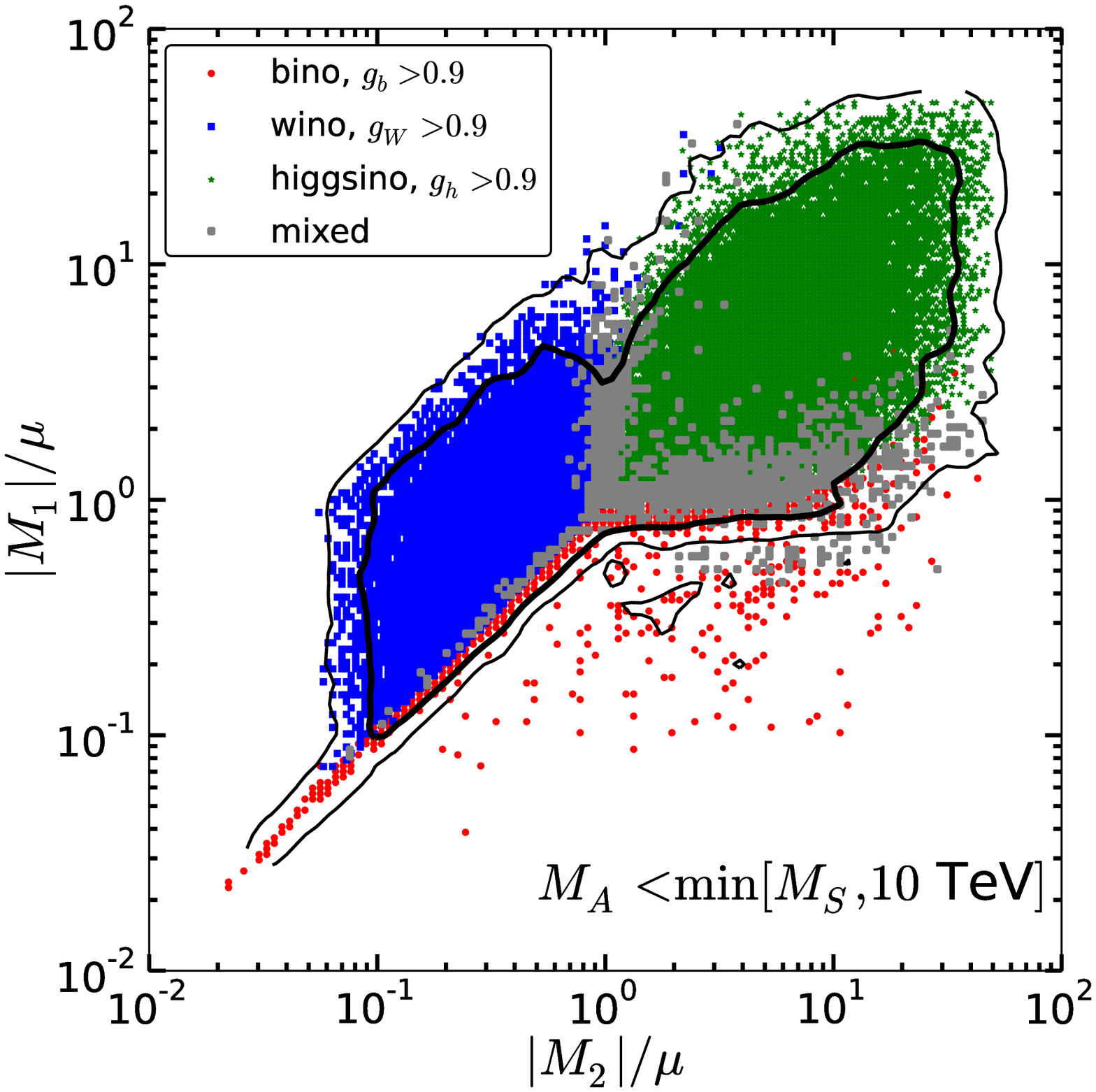}
\includegraphics[width=3.0in]{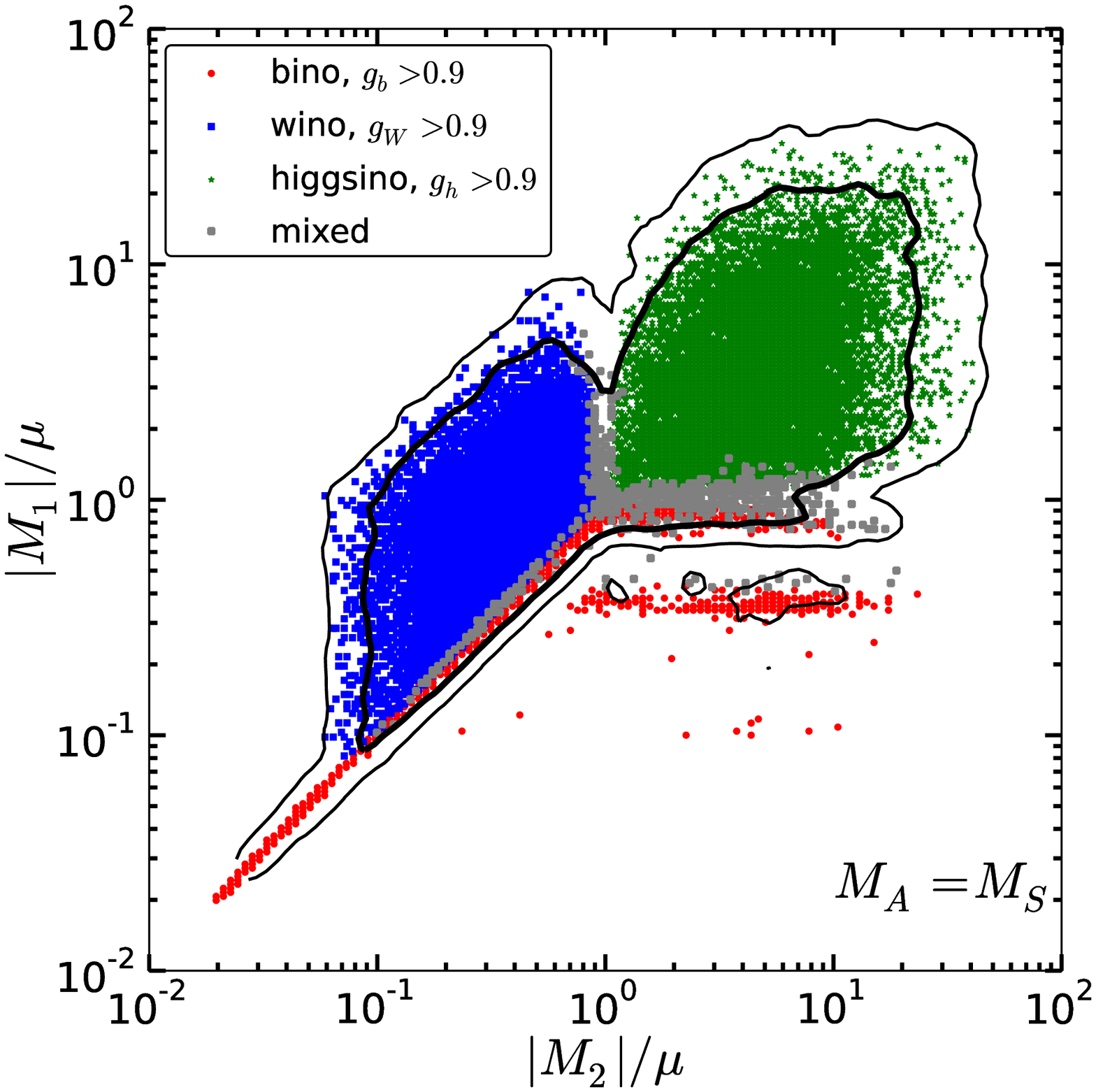}\\
\caption{\small \label{fig:inputs}
The marginalized posterior (contours) and
the profiled likelihood (scatter points) PDFs
in the ($|M_2|/\mu$, $|M_1|/\mu$) plane for
the scenarios A (left) and B (right).
%
All the three parameters are the values at the scale $M_S$.
The inner (outer) contour corresponds to $2\sigma\, (3\sigma)$ 
credible region (CR)
but the scatter points represent the $2\sigma$ profile likelihood region.
The regions with
$g_b>0.9$ (bino-like), $g_W>0.9$ (wino-like), and $g_h>0.9$ (higgsino-like)
are colored in red, blue, and green. The
gray region is for the mixed $\chi^0_1$, see the text.
}
\end{figure}

In Fig.~\ref{fig:inputs} we present the probability density functions
(PDFs) for marginalized posterior and
profiled likelihood in the ($|M_2|/\mu$, $|M_1|/\mu$) plane. All the
experimental constraints in Table~\ref{tab:exp_constraints}
are applied and we make comparisons of the scenarios
A (left) and B (right).
We represent
the bino-like, wino-like, higgsino-like and mixed neutralinos in
red, blue, green and gray, respectively.
Precisely, we identify the lightest neutralino $\chi^0_1$
as bino-, wino- or higgsino-like
when the corresponding fraction
$g_b>0.9$, $g_W>0.9$ or $g_h>0.9$, respectively.
\footnote{The parameters $g_{b,W,h}$ are defined as
$g_b =Z^{2}_{\rm{bino}}$, $g_W=Z^{2}_{\rm{wino}}$, and
$g_h=Z_{H_u}^2 + Z_{H_d}^2$ when $\chi^0_1$ is decomposed into
bino, wino, and higgsinos as follows
\[
\chi^0_1=Z_{\rm{bino}} \tilde{B} +Z_{\rm{wino}} \tilde{W}+
Z_{H_u} \tilde{H_u} + Z_{H_d} \tilde{H_d}\,.
\]}
Otherwise we identify it is the mixed lightest neutralino.
%
%
%
%
%
Comparing the scenarios A and B, we can see that the difference lies in
the bino region. This is because the bino-like $\chi^0_1$
can satisfy the relic abundance constraint only through
$Z/h$-resonance in the scenario B, where $A/H$-funnel does not 
work because $M_A=M_S \gsim 3$ TeV.
In fact, the mechanism of $Z/h$-resonance requires a small fraction of
higgsino but it cannot be too large because of the constraint
from the Fermi dSphs gamma ray measurement.
In particular, we find that the higgsino composition
is between $0.06$ to $0.1$ in the $h$ resonance region
which leads to the ratio $|M_1|/\mu \sim 0.4$.

\begin{figure}[t!]
\centering
\includegraphics[width=3.0in]{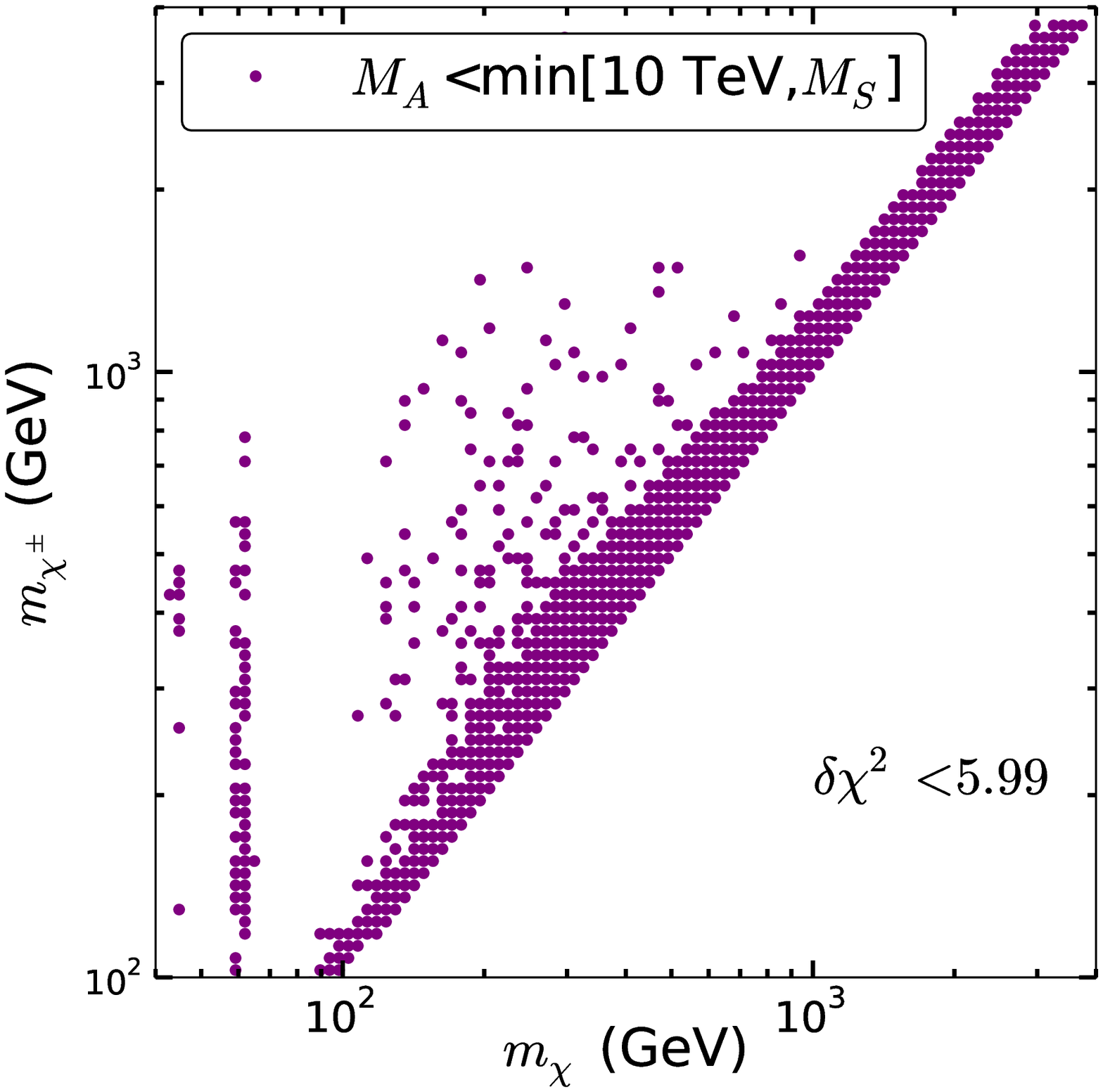}
\includegraphics[width=3.0in]{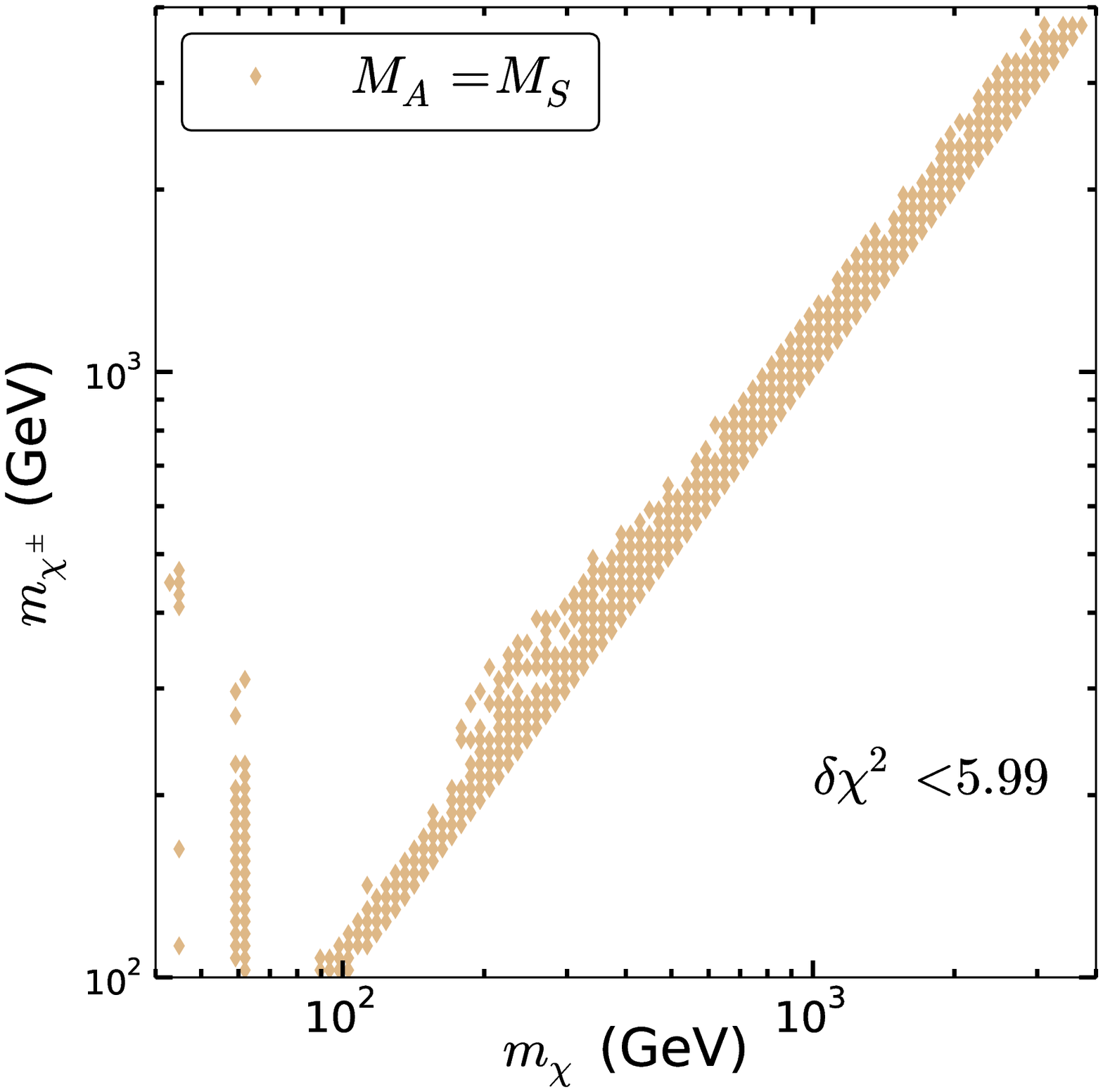}
\caption{\label{fig:mx_mxpm}
The  points with $\delta\chi^2<5.99$ scattered on the
($\mx$, $m_{\chi^\pm}$) plane for the scenario A (left) and
B (right).
}
\end{figure}

Furthermore, we find that the chargino-neutralino coannihilation
working in reducing the relic abundance in both scenarios.
Being different from the original split SUSY framework (scenario B),
one can obtain the correct relic abundance in scenario A without
resorting to the coannihilation mechanism thanks to the
intermediate Higgses $A$ and $H$.
To address this point, we show in Fig.~\ref{fig:mx_mxpm}
the points with $\delta\chi^2<5.99$ on the
($\mx$, $m_{\chi^\pm}$) plane for the scenario A (left) and
B (right).
In addition to the $Z/h$-resonance regions
around $m_{\chi^0_1} \sim  50\,, 60$ GeV and
the chargino-neutralino coannihilation region along
the $m_{\chi^0_1} = m_{\chi^\pm}$ line,
we observe there are more points appearing in the scenarios A (left
panel) due to the $A/H$-funnel.
We find that the $A/H$-funnel region disappears when
$m_{\chi^0_1}>1$ TeV, because the decay widths of $A$ and $H$ become too large
and the Breit-Wigner resonance effect is not strong enough to
reduce the relic abundance when $M_{H,A}\gsim 2$ TeV.

\begin{figure}[t!]
\centering
\includegraphics[width=3.0in]{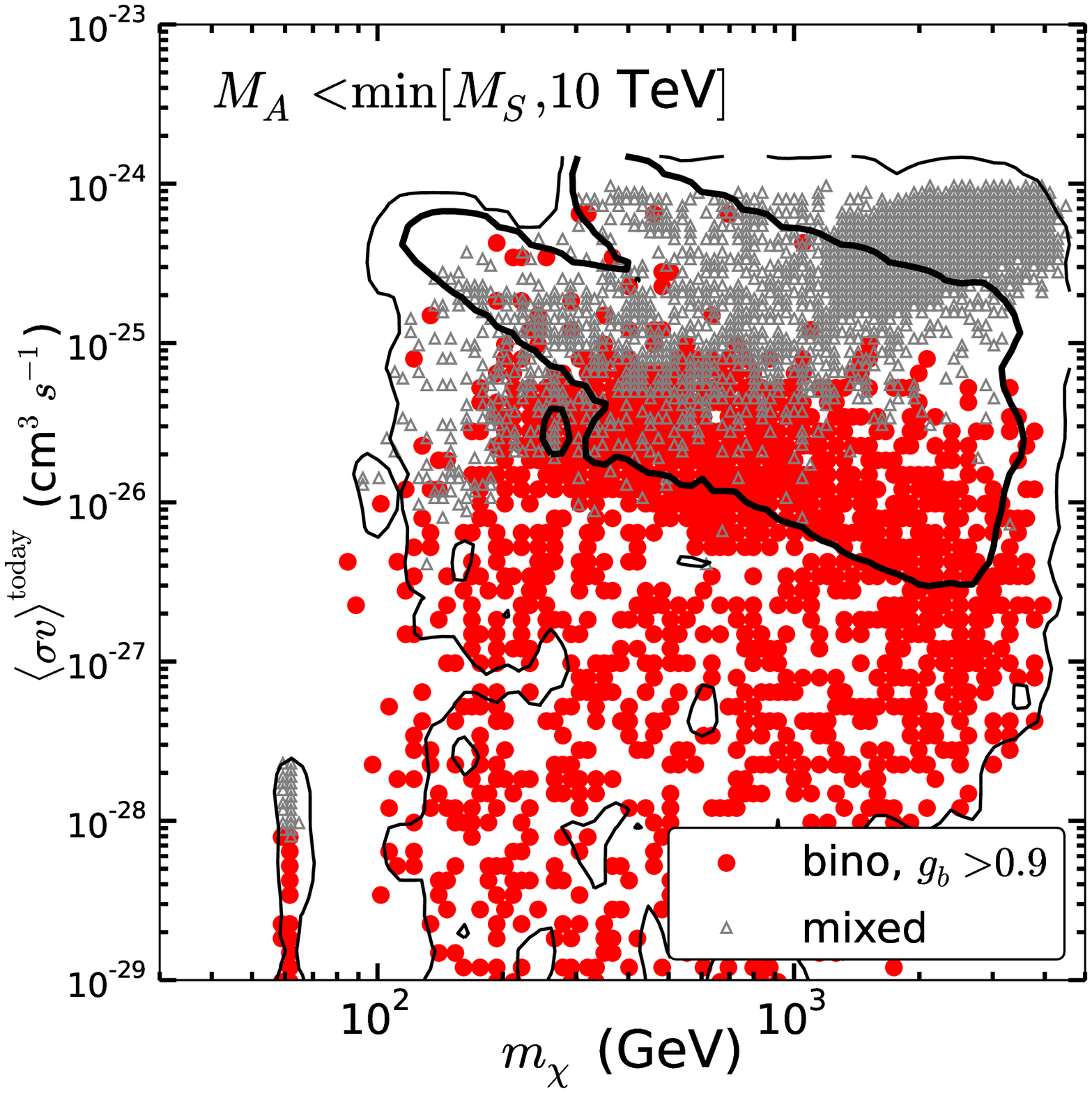}
\includegraphics[width=3.0in]{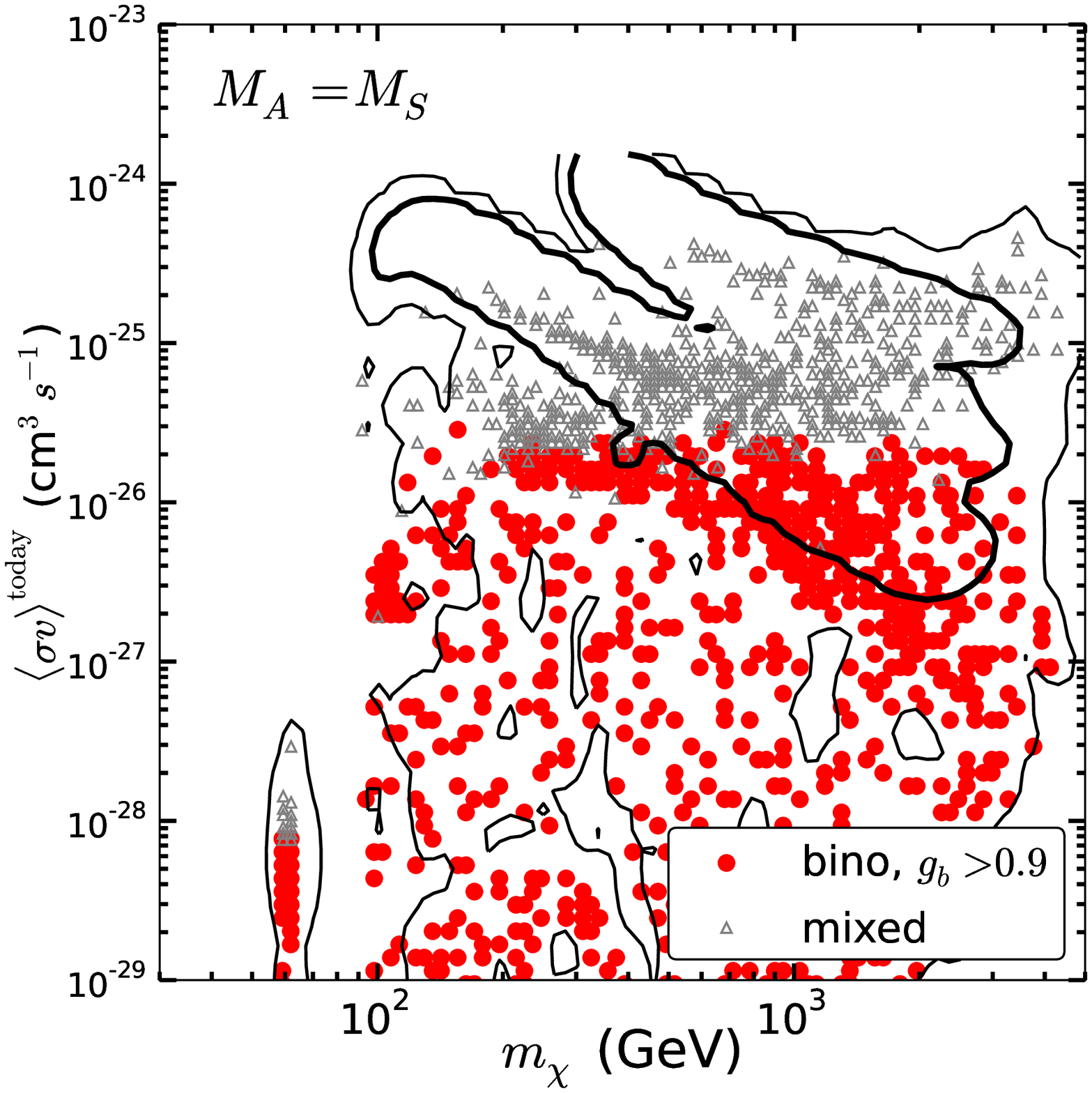}\\
\includegraphics[width=3.0in]{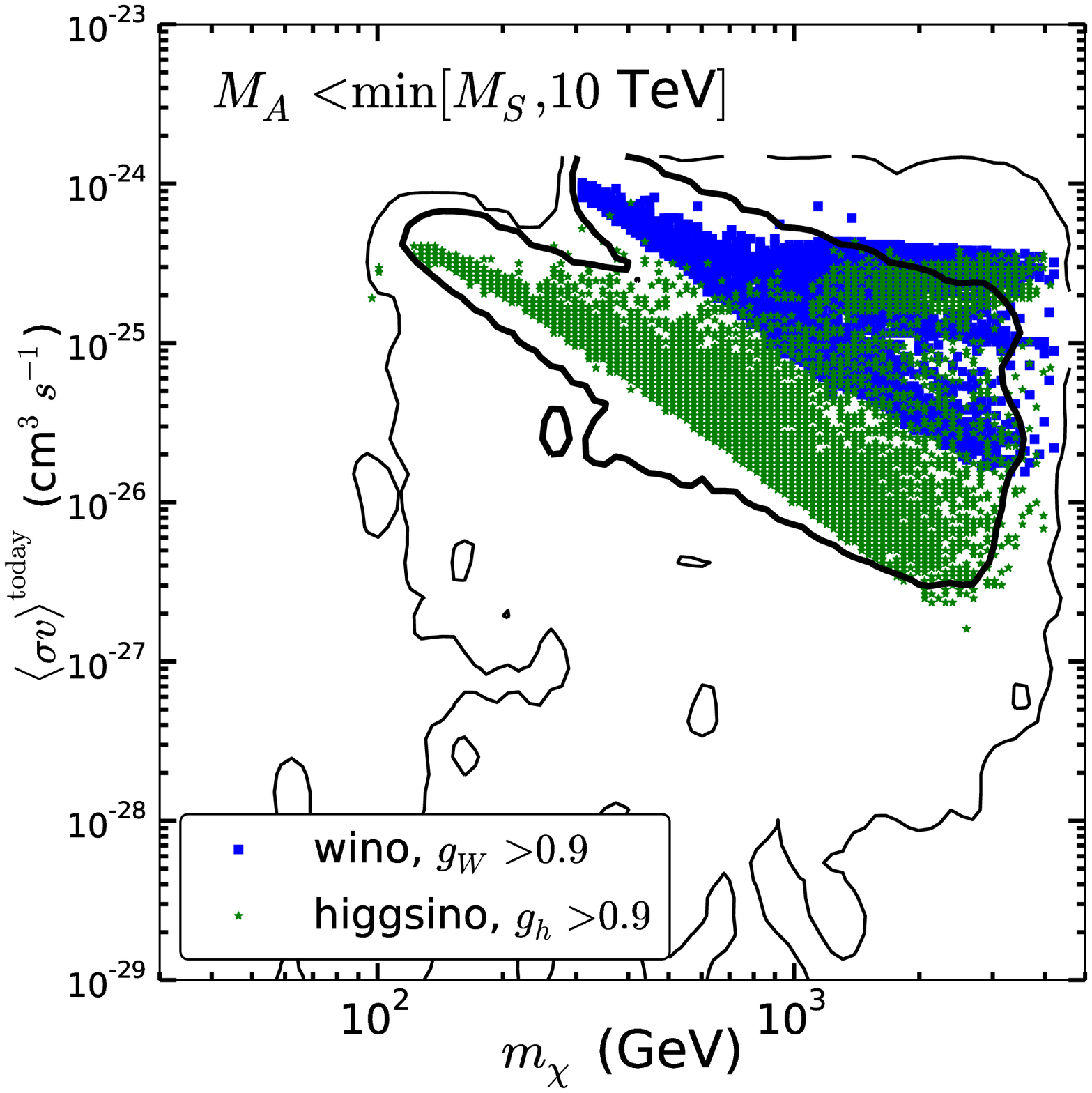}
\includegraphics[width=3.0in]{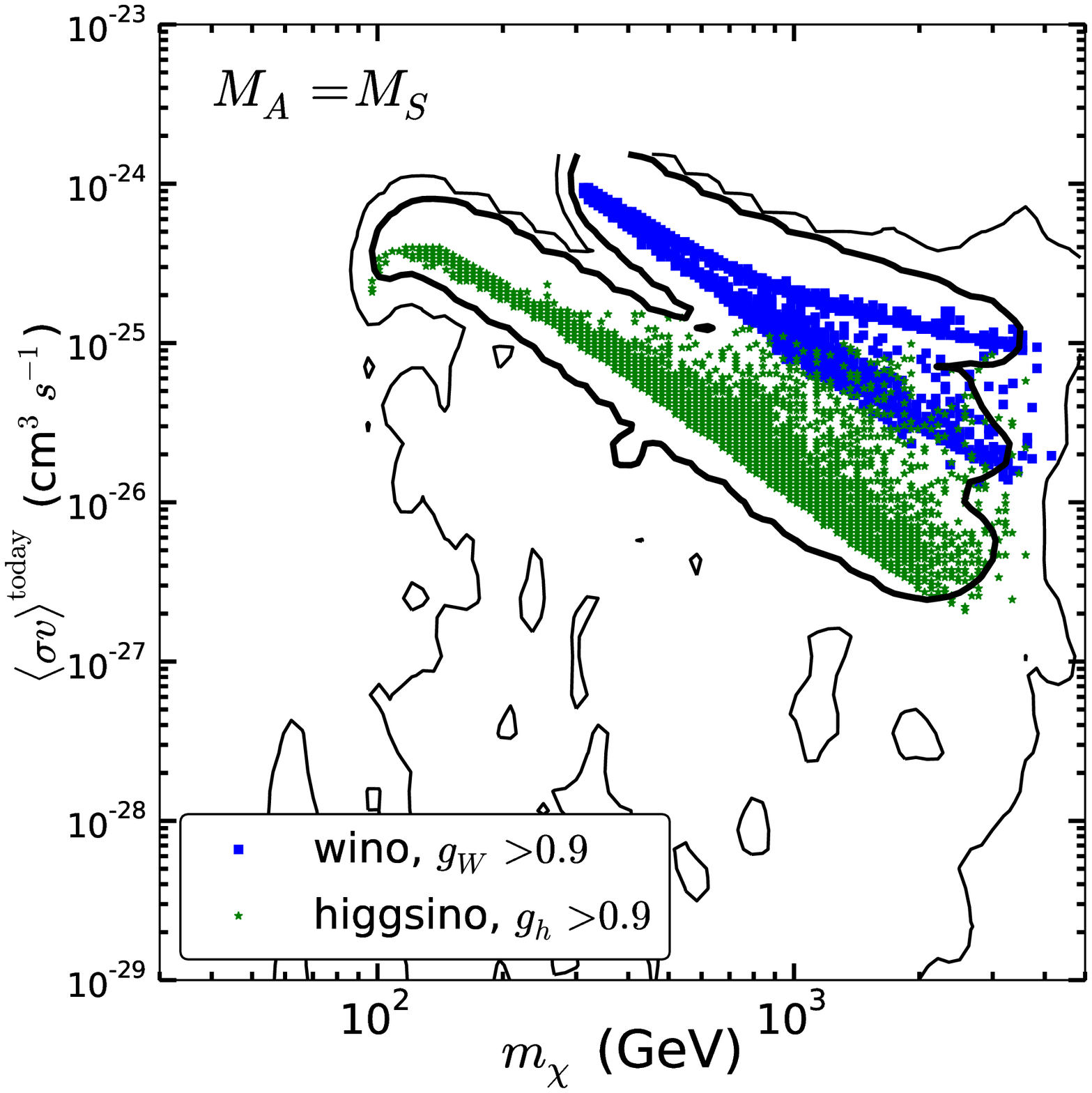}\\
\caption{\small \label{fig:mx_sigmav}
Marginalized posterior PDF (contours) and
profiled likelihood PDF (scatter points)
in the ($\mx$, $\sigmav$) plane for the scenario A (left) and B (right).
The inner (outer) contours bounded the $2(3)$-$\sigma$ CR.
All the scatter points superimposing on the contours agree
with likelihood in the criteria $\delta\chi^2<5.99$.
The red dots, blue squares, green stars, and gray triangle
are for the bino-like, wino-like, higgsino-like, and mixed neutralino,
respectively.
}
\end{figure}

In Fig.~\ref{fig:mx_sigmav},
we show the marginalized 2D posterior
$2$- and $3$-$\sigma$ credible regions (CRs) for the scenario A (left)
and B (right) in the ($\mx$, $\sigmav$) plane.
We also show the PL 2-$\sigma$ region (scattered points)
for the bino-like (red) and mixed (gray) $\chi^0_1$ in the upper frames
and the wino-like (blue) and higgsino-like (gray) $\chi^0_1$ 
in the lower frames.
Here $\sigmav$ denotes the annihilation cross section at the present time
which is relevant to the DM indirect detections and
through which one may easily identify different
mechanisms for the relic abundance.

When $\mx<100\gev$, via the $Z/h$ resonances,
the marginalized posterior CRs are located
at the bino-like neutralino region
with a small amount of higgsino component
in both scenarios (see the upper frames).
Although the $Z/h$-resonance channels have very good likelihoods,
they only fall into the $3\sigma$ (99.73\%) CR owing to
the small prior volume effect.
The similar effect happens for the bino-like $\chi^0_1$
when $m_{\chi^0_1}>100$ GeV and the correct relic abundance is obtained
by the $\chi\chi^\pm$ coannihilation.
The fact that more parameter space survives in the scenario A (left)
than scenario B (right) is due to the $A/H$-funnel.
Nevertheless, most of the additional parameter space is a result of
the mixture mechanism between $A/H$-funnel and coannihilation.
In the lower frames, we observe that the $2\sigma$ CR has the
wino-like branch (blue) with the higher $\sigmav$ than 
the higgsino-like one (green).
For the wino-like branch, the relic abundance is mainly reduced by
the wino-like DM annihilation into $W^+W^-$ pairs.
However when $\mx\gsim 3$ TeV, the wino DM
cannot give the correct relic abundance as is well known.
This mass limit can be slightly extended
if coannihilation is taken into account.
Since the wino DM have higher annihilation
cross sections, the indirect detection constraint is stringent.
Indeed, the lower bound for the wino-like neutralino mass is about 
$300\gev$ from
the Fermi dSphs gamma ray constraints.
Incidentally, the lower bound for the higgsino-like neutralino mass is
about $100\gev$, set by the LEP limit of $m_{\chi_1^\pm}>103.5$.
We further see there is no particular lower bound for the
bino-like or mixed neutralino, as seen from the upper frames.

\begin{figure}[t!]
\centering
\includegraphics[width=3.0in]{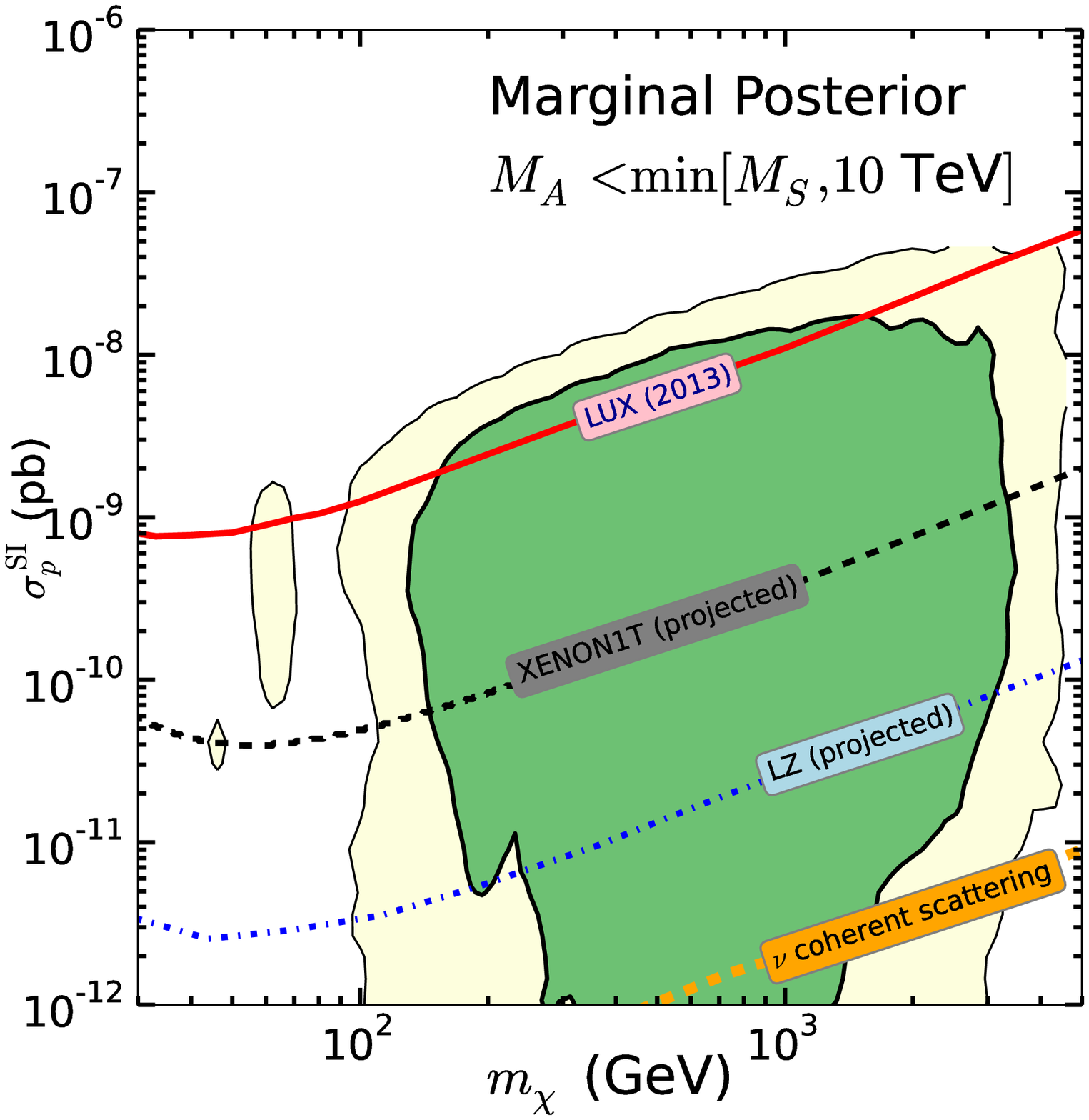}
\includegraphics[width=3.0in]{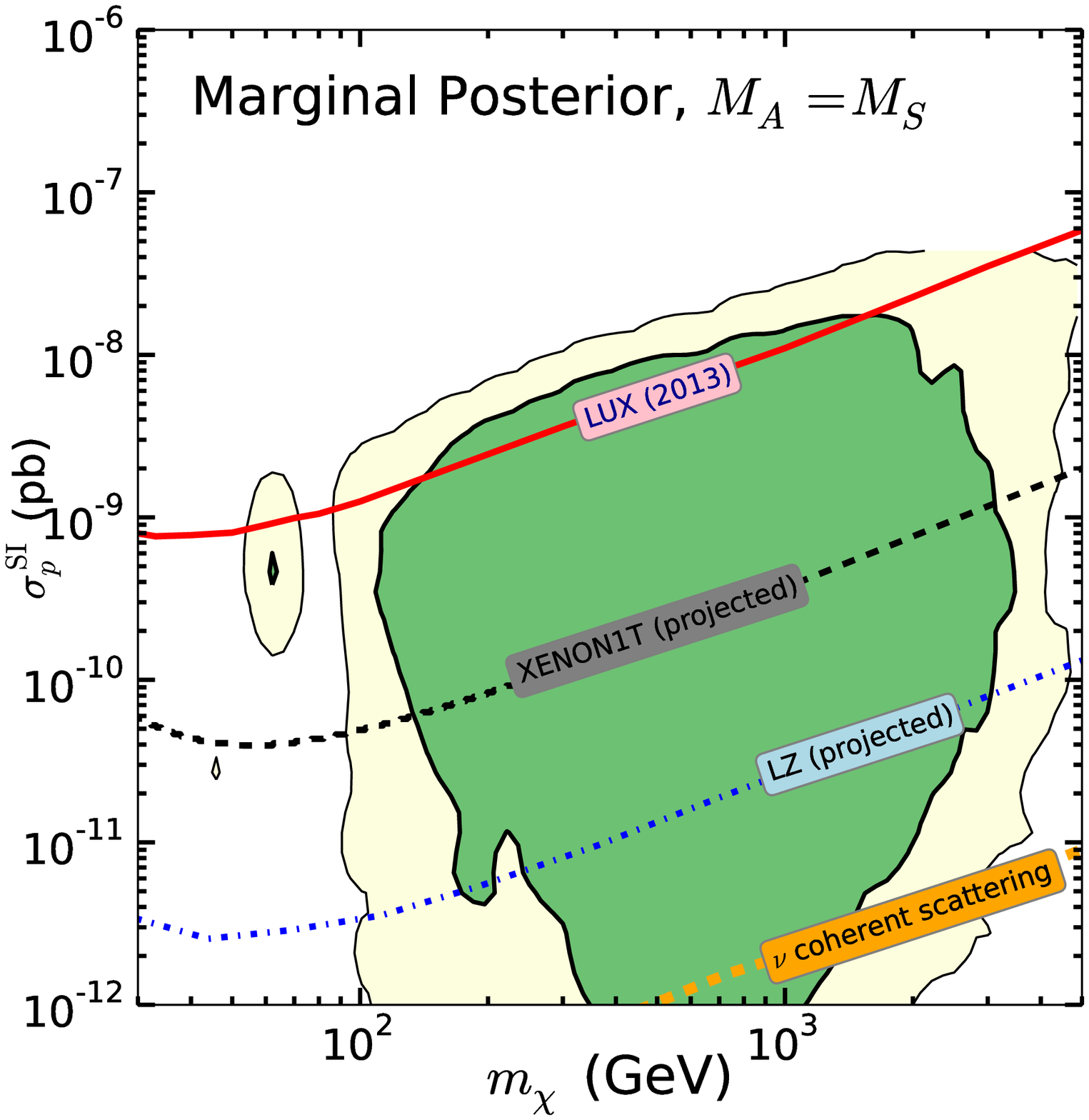}\\
\caption{\small \label{fig:mx_sigsip}
The marginal posterior for the 95\% and 99.73\% CRs
in ($\mx$, $\sigsip$) plane.
The left (right) panel is for the scenario A (B).
}
\end{figure}

Finally, in Fig.~\ref{fig:mx_sigsip} we show the marginalized 2D posterior PDF
$2\sigma$ and $3\sigma$ contours in the ($\mx$, $\sigsip$) plane.
The red solid line denotes the recent LUX result, the black dashed line
the XENON1T  projected sensitivity, and the blue dash-dotted line the
LZ projected sensitivity~\cite{Malling:2011va}.
The orange dashed line represents the approximate line below which
the DM signal becomes hardly distinguishable from
the signals from the coherent scattering
of the $^{8}B$ solar neutrinos, atmospheric neutrinos and
diffuse supernova neutrinos with nuclei.
We observe that a part of $2$-$\sigma$ CR is below
the LZ projected sensitivity.
We can see that, in the $2$-$\sigma$ CRs, there is no significant
difference between the scenarios A and B. The $3$-$\sigma$ CRs are
slightly different in the lower $\sigsip$ region.
Moreover, in both scenarios, the future 7-tons experiments,
LZ, can set a lower limit on the neutralino DM at $\mx>100\gev$.

\section{Discussion}
In this work, we have studied a ``modified split SUSY'' scenario, characterized
by two separate scales -- the SUSY-breaking scale $M_S$ and the heavy
Higgs-boson mass scale $M_A$.  This is different from the split SUSY
scenario, in which the scale $M_A$ is also set at $M_S$.  The
current scenario is motivated by
(i) the absence of direct SUSY signals from the searches of scalar quarks
up to a few TeV,
(ii) the observed Higgs boson is somewhat on the heavy side which needs a
large radiative correction to the tree-level mass from heavy stops, and
(iii) absence of signals from heavy Higgs bosons $A/H$ and $H^\pm$
which can be as light as a few hundred GeV.
Therefore, the choice of $M_A$ need not be as large as $M_S$. We have studied
two scenarios: (i) $M_A \le {\rm min}(M_S, {\rm 10}\, {\rm Tev})$
and (ii) $M_A = M_S$ (the same as split SUSY).

If both $M_A$ and $M_S$ are set equal with $M_A<10$ TeV, as shown
in Fig.~\ref{fig:mA_mS}, only a small region with $M_S \lsim 10^4$ GeV 
and large $\tan\beta$ is allowed.
Nevertheless, if $M_A$ and $M_S$ are set at different values,
much larger parameter space with a wide range of $\tan\beta$ is allowed.
With more parameter space we have performed a careful analysis using
all dark matter constraints and collider limits.

Because of two distinct scales $M_S$ and $M_A$ the running of the
soft parameters and couplings are separated in two steps. We start
with the set of RGEs given in appendix A to run from $M_S$ down to
$M_A$ and perform the matching at the scale $M_A$. Then run from
$M_A$ down to the electro-weakino scale $M_\chi \equiv \sqrt{\mu \times M_2}$
with the set of RGEs of split SUSY. Because of this two-step RGEs
the predictions for DM observables and the Higgs boson mass are
more reliable than just a single-step RGE.

We have scanned the MSSM parameter space characterized by the two scales:
$M_S$ and $M_A$ subjected to many existing experimental constraints:
invisible widths of the $Z$ boson and the Higgs boson, the chargino
mass limit, relic abundance of the LSP, spin-independent cross sections
from direct detection, and the gamma-ray data from indirect detection.
We found interesting survival regions of parameter space with features
of either chargino-neutralino coannihilation, the $A/H$ funnel, or
wino-like. These regions survive because of the large enough annihilation
to reduce the relic abundance to the observed values, as well
as give a large enough Higgs boson mass to fit to the observed value.
Finally,
the survived parameter space can be further scrutinized by near future
direct detection experiments such as XENON1T and LZ.

We offer a few important comments as follows.
\begin{enumerate}
\item We used the Higgs boson mass in the range
range $121.1 < m_h < 129.1$ GeV to search for suitable $M_S$.
Since $m_h$ is on the rather
heavy side, it requires a large radiative correction to the tree-level
mass. This can be achieved by a large stop mass and/or large mixing in the stop
sector. Since the radiative correction is proportional to some powers of
$\tan\beta$, a smaller $\tan\beta$ requires then a larger $M_S$ in order
to achieve a large enough $m_h$. Typically, $M_S \gsim 10^{5-6}$ GeV for
$\tan\beta < 3$.  For large enough $\tan\beta$ the values
of $M_S$ is more or less independent of $M_A$.

\item On the other hand, if we set $M_A = M_S$ as we do in scenario A,
the allowed $M_S$ is rather short from about $10^3 - 10^4 $ GeV with
large $\tan\beta$ (see Fig.~\ref{fig:mA_mS}).

\item An interesting region that satisfies the relic abundance constraint
is characterized by nearly degenerate mass among the first two
neutralinos and the lightest chargino, indicated by $M_2/\mu \approx
M_1/\mu \approx 1$. The increased effective annihilation cross section
can help reducing the relic abundance.

\item Another interesting region is the $Z/h$ resonance region
($m_{\chi^0_1} \sim 50-60$ GeV), though 
it is relatively fine-tuned region because
of the narrow width of the $Z$ boson and the Higgs boson.

\item Yet, another interesting survival region is the $A/H$ funnel region.
If $m_{\chi^0_1}$ falls around the vicinity of $m_{A/H}/2$ the resonance effect
is strong, provided that the width is not too large.  This can be achieved
for $m_{\chi^0_1} \lsim  1$ TeV, that is $M_{A/H} \lsim 2$ TeV.  In scenario B,
where $M_S = M_A$, large values of $M_S$ then cannot be accepted because
the $A/H$ funnel is not working efficiently.
However, in scenario A, where $M_A < M_S$, the $A/H$ funnel can be
very effective in reducing the relic abundance, thus more parameter space
is allowed.

\item Both wino-like and higgsino-like LSPs have large annihilation
cross sections. The allowed mass for $m_{\chi^0_1}$ ranges from
about 300 GeV to 3 TeV for wino-like LSP while from about 100 GeV to
2 TeV for higgsino-like LSP.

\item The current allowed parameter space has a large region
below the current LUX limit $\sigma^{\rm SI}_p \lsim 10^{-9}$ pb.
Although the future XENON1T can improve the limit
by an order of magnitude, there is still a sizable region below the XENON1T
sensitivity. Yet, there still exist some allowed regions even with
the future 7-tons size direct detection experiment LZ.
Therefore, this modified split SUSY scenario is hard to be
excluded in the future.

\item We have used both the methods of profile likelihood and
marginal posterior.  Though these two statistical approaches have
very different methodology, the resulting 2- and 3-$\sigma$ regions
are quite consistent, as shown in the figures.

\end{enumerate}

\section*{Acknowledgment}
R.H.~is grateful to Carlos.~E.M.~Wagner, Stephen P.~Martin and 
Alessandro Strumia for useful discussions.
K.C. was supported by the National Science
Council of Taiwan under Grants No. NSC 102-2112-M-007-015-MY3.
J.S.L. was supported by
the National Research Foundation of Korea (NRF) grant
(No. 2013R1A2A2A01015406) and by Chonnam National University, 2012.
R.H. and Y.S.T. were supported by World Premier International Research 
Center Initiative (WPI), MEXT, Japan.

\begin{center}
{\Large \bf Appendix}
\end{center}

\def\theequation{\Alph{section}.\arabic{equation}}
\begin{appendix}
\setcounter{equation}{0}
\section{RGEs from $M_S$ to $M_A$}
\label{sec:rge_ours}
Here we present the one-loop
RGEs governing the running of couplings from the high SUSY scale $M_S$ to
the intermediate Higgs mass scale $M_A$.

We write the RGE for each coupling $g_i$ present in the theory,
in the $\overline{\rm MS}$ or $\overline{\rm DR}$ scheme 
(the same up to one-loop level), as
\begin{equation}
\frac{dg_i}{d\ln Q} = \frac{\beta_1(g_i)}{(4\pi)^2} .
\end{equation}
The relevant coupling constants $g_i$ include the gauge couplings
($g_s,g,g'$)
, the gaugino couplings ($\tilde{g}_d',\tilde{g}_u'$, $\tilde{g}_d,\tilde{g}_u$), the
third-generation Yukawa couplings ($y_t,y_b,y_\tau$), and the Higgs quartic
($\lambda_1,\lambda_2,\lambda_3,\lambda_4,\lambda_5,\lambda_6,\lambda_7$).

At one loop the $\beta$ functions of gauge couplings below the SUSY scale are given by
\begin{equation}
\beta_1(g_s) =-5 g_s^3,\qquad\qquad
\beta_1(g) =-1 g^3,\qquad\qquad
\beta_1(g') =\frac{23}{3}g'^3.
\end{equation}
The $\beta$ functions of gauge couplings defined by the fermion-scalar-gaugino
interaction below the SUSY scale are given by
\begin{eqnarray}
\beta _1\left(\tilde{g}_u\right) &=&
\tilde{g}_u\left(
-\frac{33}{4}g^2 - \frac{3}{4}g'^2
+\frac{11}{4}\tilde{g}_u^2+\frac{1}{2}\tilde{g}_d^2
+\frac{3}{4}\tilde{g}_u'^2
+3 y_t^2
\right)\\
\beta _1\left(\tilde{g}_d\right) &=&
\tilde{g}_d\left(
-\frac{33}{4}g^2 - \frac{3}{4}g'^2
+\frac{11}{4}\tilde{g}_d^2+\frac{1}{2}\tilde{g}_u^2
+\frac{3}{4}\tilde{g}_d'^2
+3 y_b^2+y_\tau^2
\right) \\
\beta _1\left(\tilde{g}_u'\right) &=&
\tilde{g}_u'\left(
-\frac{9}{4}g^2 - \frac{3}{4}g'^2
+\frac{9}{4}\tilde{g}_u^2
+\frac{5}{4}\tilde{g}_u'^2+\frac{1}{2}\tilde{g}_d'^2
+3 y_t^2
\right) \\
\beta _1\left(\tilde{g}_d'\right) &=&
\tilde{g}_d' \left(
-\frac{9}{4}g^2 - \frac{3}{4}g'^2
+\frac{9}{4}\tilde{g}_d^2
+\frac{5}{4}\tilde{g}_d'^2+\frac{1}{2}\tilde{g}_u'^2
+3 y_b^2+y_\tau^2
\right)
\end{eqnarray}

The $\beta$ functions of 3rd generation Yukawa interactions below the SUSY scale are
given by
\begin{eqnarray}
\beta _1\left(y_t\right) &=& y_t \left(
\frac{9}{2} y_t^2 + \frac{1}{2} y_b^2
-8 g_s^2 -\frac{9}{4} g^2 -\frac{17}{12} g'^2
+\frac{3}{2}\tilde{g}_u^2
+\frac{1}{2}\tilde{g}_u'^2
 \right) \\
\beta _1\left(y_b\right) &=&y_b\left(
\frac{9}{2} y_b^2 + \frac{1}{2} y_t^2 + y_\tau^2
-8 g_s^2 -\frac{9}{4} g^2 -\frac{5}{12} g'^2
+\frac{3}{2}\tilde{g}_d^2
+\frac{1}{2}\tilde{g}_d'^2
\right) \\
\beta _1\left(y_\tau\right) &=&y_\tau \left(
\frac{5}{2} y_\tau^2 + 3 y_b^2
-\frac{9}{4} g^2 - \frac{15}{4} g'^2
+\frac{3}{2}\tilde{g}_d^2
+\frac{1}{2}\tilde{g}_d'^2
\right)
\end{eqnarray}

The $\beta$ functions of Higgs quartic couplings defined by
Haber and Hempfling \cite{Haber:1993an} are given by
\begin{eqnarray}
\beta_1\left(\lambda_1\right) &=& \bigg(
24\lambda_1^2 + 4\lambda_3^2 + 4(\lambda_3+\lambda_4)^2 + 4\lambda_5^2 + 48\lambda_6^2
\nonumber \\
& & + \frac{3}{8}\Big(2g^4 + (g^2+g'^2)^2\Big) - \Big(2\tilde{g}_d^4 +
\frac{1}{2}(\tilde{g}_d^2 + \tilde{g}_d'^2)^2\Big) - 2N_cy_b^4 - 2y_\tau^4 \nonumber \\
& & + 4\lambda_1\gamma_d \bigg), \\
\beta_1\left(\lambda_2\right) &=& \bigg(
24\lambda_2^2 + 4\lambda_3^2 + 4(\lambda_3+\lambda_4)^2 + 4\lambda_5^2 + 48\lambda_7^2
\nonumber \\
& & + \frac{3}{8}\Big(2g^4 + (g^2+g'^2)^2\Big) - \Big(2\tilde{g}_u^4 +
\frac{1}{2}(\tilde{g}_u^2 + \tilde{g}_u'^2)^2\Big) - 2N_cy_t^4 \nonumber \\
& & + 4\lambda_2\gamma_u \bigg), \\
\beta_1\left(\lambda_3\right) &=& \bigg(
(\lambda_1+\lambda_2)(3\lambda_3+\lambda_4) + 8\lambda_3^2 + 4\lambda_4^2 +
4\lambda_5^2 + 8\lambda_6^2 + 8\lambda_7^2 + 32\lambda_6\lambda_7 \nonumber \\
& & + \frac{3}{8}\Big(2g^4 + (g^2-g'^2)^2\Big) -
\Big(2\tilde{g}_u^2\tilde{g}_d^2 +
\frac{1}{2}(\tilde{g}_u^2-\tilde{g}_u'^2)(\tilde{g}_d^2-\tilde{g}_d'^2)\Big)
- 2N_cy_b^2y_t^2 \nonumber \\
& & + \lambda_3(2\gamma_d+2\gamma_u) \bigg), \\
\beta_1\left(\lambda_4\right) &=& \bigg(
4\lambda_4(\lambda_1+\lambda_2+4\lambda_3+2\lambda_4) + 16\lambda_5^2 + 20\lambda_6^2 +
20\lambda_7^2 + 8\lambda_6\lambda_7 \nonumber \\
& & + \frac{3}{2}g^2g'^2 + 2\tilde{g}_d^2\tilde{g}_u^2 - \tilde{g}_d^2\tilde{g}_u'^2 -
\tilde{g}_d'^2\tilde{g}_u^2 + 2N_cy_b^2y_t^2 \nonumber \\
& & + \lambda_4(2\gamma_d+2\gamma_u) \bigg), \\
\beta_1\left(\lambda_5\right) &=& \bigg(
4\lambda_5(\lambda_1+\lambda_2+4\lambda_3+6\lambda_4) + 20(\lambda_6^2 + \lambda_7^2) +
8\lambda_6\lambda_7 \nonumber \\
& & + \lambda_5(2\gamma_d+2\gamma_u) \bigg), \\
\beta_1\left(\lambda_6\right) &=& \bigg(
4\lambda_6(6\lambda_1+3\lambda_3+4\lambda_4+5\lambda_5) +
4\lambda_7(3\lambda_3+2\lambda_4+\lambda_5) \nonumber \\
& & + \lambda_6(3\gamma_d+\gamma_u)\bigg), \\
\beta_1\left(\lambda_7\right) &=& \bigg(
4\lambda_7(6\lambda_2+3\lambda_3+4\lambda_4+5\lambda_5) +
4\lambda_6(3\lambda_3+2\lambda_4+\lambda_5) \nonumber \\
& & + \lambda_7(\gamma_d+3\gamma_u)\bigg),
\end{eqnarray}
where
\begin{eqnarray}
\gamma_d &=& N_cy_b^2 + y_\tau^2 - \frac{3}{4}(3g^2 + g'^2) +
\frac{1}{2}(3\tilde{g}_d^2 + \tilde{g}_d'^2), \\
\gamma_u &=& N_cy_t^2 - \frac{3}{4}(3g^2 + g'^2) + \frac{1}{2}(3\tilde{g}_u^2 +
\tilde{g}_u'^2).
\end{eqnarray}

The $\beta$ functions of gaugino mass parameters and the SUSY $\mu$ term are given by
\begin{eqnarray}
\beta_1\left(M_3\right) &=& -18g_s^2M_3  \\
\beta_1\left(M_2\right) &=& (-12g^2+\tilde{g}_u^2+\tilde{g}_d^2)M_2  \\
\beta_1\left(M_1\right) &=& (\tilde{g}_u'^2+\tilde{g}_d'^2)M_1 \\
\beta_1\left(\mu \right) &=&
\left(-\frac{9}{2}g^2+\frac{3}{4}\tilde{g}_u^2+\frac{3}{4}\tilde{g}_d^2-\frac{3}{2}g'^2+\frac{1}{4}\tilde{g}_u'^2+\frac{1}{4}\tilde{g}_d'^2\right)\mu
\end{eqnarray}

\setcounter{equation}{0}
\section{The statistical framework}
\label{sec:stat}
To calculate 
 the probability of MSSM parameter given the experimental data,
one can employ Bayes's' theorem to compute the posterior probability density 
function,
\begin{equation}
\label{eq:Bayes}
p(\theta,\phi|d)=\frac{\mathcal{L}(d|\theta,\phi)\pi(\theta,\phi)}{\mathcal{Z}(d)}.
\end{equation}
Here, we denote the MSSM parameters and DM direct detection nuisance parameters
as $\theta$ and $\phi$, respectively. The likelihood 
$\mathcal{L}(d|\theta,\phi)$
is the probability of obtaining experimental data for observables given the MSSM
parameters. The prior knowledge of MSSM parameter space is presented as
prior distribution $\pi(\theta,\phi)$.
Our MSSM prior ranges and distributions are tabulated in Table \ref{tab:inputs}.
Finally, the evidence of the model in the denominator
can be merely a normalization factor, 
because we are not interested in model comparison.

The Bayesian approach allows us to simply get ride of the unwanted parameters
by using marginalization. For example, if there would be $n$ 
free model parameters, $r_{i=1,...,n}$,
but one is only interesting in the two-dimensional figure ($r_1$, $r_2$), 
the marginalization can be written as
\begin{equation}
p(r_1,r_2|d)=\int p(r_1,...,r_n|d) \prod_{i=3}^{n}dr_i.
\end{equation}
An analogous procedure can be performed with the observables.
One should keep in mind that a poor prior knowledge or likelihood
function can raise a volume effect. In other words,
some regions gain more weight from higher prior probability but
fine-tuning regions such as resonance regions for relic abundance likelihood
only have lower prior probability.
Although this is the feature of Bayesian statistics,
in order to manifest these fine-tuning regions,
we still present both profile likelihood and marginal posterior
method at the same time.

In Bayesian statistics, a credible region (CR) is the smallest region,
$\mathcal{R}$, in the best agreement with experiments bounded
with the fraction $\varrho$
of the total probabilities. For example at MSSM ($M_1$, $M_2$) plane,
the $\varrho$ credible region can be written as
 \begin{equation}
\frac{\int_{\mathcal{R}}p(M_1,M_2|d)dM_1 dM_2}
{\rm{normalization}}
=\varrho,
\end{equation}
where the normalization in the denominator is
the total probability with $\mathcal{R}\to\infty$.
In this paper, we have shown $\varrho=0.95$ and $\varrho=0.9973$
corresponding to $2\sigma$ and $3\sigma$ credible region.
As the comparison, we also present the scatter points
with selected criteria $\delta\chi^2=-2\ln\mathcal{L/L_{\rm{max}}}
\le 5.99$.
This criteria is $2\sigma$ confidence region of Profile Likelihood method
in 2 degrees of freedom.
We can see from our result that most of $2\sigma$ confidence region of
PL method is similar to the 3 $\sigma$ credible region
in MP method. We would like to note that
the total profile likelihood here
takes the likelihoods including 
the nuisance parameters distribution,
which is the prior distribution in marginal posterior method.

%

\end{appendix}


\end{document}